\documentclass[preprint,prb,eqsecnum,preprintnumbers,amsmath,amssymb]{revtex4}

\usepackage{graphicx}
\usepackage{dcolumn}
\usepackage{bm}
\def\lesssim{\mathrel{\mathpalette\vereq<}}
\def\vereq#1#2{\lower3pt\vbox{\baselineskip1.5pt \lineskip1.5pt
\ialign{$#1\hfill##\hfil$\crcr#2\crcr\sim\crcr}}}

\def\alt{\lesssim}

\def\vergt#1#2{\lower5pt\vbox{\baselineskip1.5pt \lineskip-1.0pt
\ialign{$#1\hfill##\hfil$\crcr#2\crcr<\crcr}}}

\def\citeonline#1{\onlinecite{#1}}

\begin{document}
\title{%
Crossover between Local-Moment Magnetism and \\
Itinerant-Electron Magnetism  in the $t$-$J$ Model 
}
\author{Fusayoshi J. O{\scriptsize HKAWA}\thanks{fohkawa@phys.sci.hokudai.ac.jp}}
\affiliation{Division of Physics, Graduate School of Science, 
Hokkaido University, Sapporo 060-0810, Japan}
\date{\today}
%
\begin{abstract}  
A Kondo-lattice theory is applied to the $t$-$J$ model on a quasi-two dimensional
lattice.  The Kondo temperature $T_{\rm K}$ or $k_{\rm B}T_{\rm K}$ is defined as
an energy scale of local quantum spin fluctuations, with $k_{\rm B}$ the Boltzmann
constant. 
 The bandwidth $W^*$ of quasiparticles, which is approximately given by
$W^*\simeq 4 k_{\rm B}T_{\rm K}$, is renormalized by the Fock-type term of the
superexchange interaction. When  the lifetime width $\gamma$ of quasiparticles is
so small and temperature  $T$ is so low  that $\gamma/W^*\alt 1$ and
$k_{\rm B}T/W^*\alt 1$, the renormalization is large.
Even in the limit of the  half filling, for example, 
the bandwidth $W^*$ is nonzero and of the order of
$|J|$, with $J$ the superexchange
interaction constant between nearest neighbors. 
The Kondo temperature $k_{\rm B}T_{\rm K}$, which also gives a measure of the strength 
of the quenching of magnetic moments by local quantum spin fluctuations,
increases away from the half filing; it is smaller for larger $\gamma$.  
Therefore, local-moment magnetism, which is characterized by
$T_{\rm N}\gg T_{\rm K}$,  with $T_{\rm N}$ the N\'{e}el temperature, appears for
almost half fillings, and itinerant-electron magnetism, which is characterized by
$T_{\rm N}\alt T_{\rm K}$, appears for fillings away from the half filling; 
$T_{\rm N}$ is higher   for larger $\gamma$.
The antiferromagnetic region as a function of electron fillings is wider 
for larger $\gamma$; this result implies that it  is wider in more
disordered systems than it is in less disordered ones.
The difference or asymmetry of  disorder between
electron-doped and hole-doped cuprate oxide superconductors 
must be, at least partly, responsible for that
of antiferromagnetic phases between them.
\end{abstract}  
\pacs{71.30.+h, 75.30.Kz, 71.10.-w, 75.10.Lp}

\maketitle

\section{Introduction}\label{SecIntroduction}

The discovery\cite{bednorz} in 1986 of high critical-temperature 
(high-$T_{\rm c}$) superconductivity 
in  cuprate oxides has revived intensive and extensive
studies on strong electron correlations because it occurs in the vicinity of the 
Mott-Hubbard metal-insulator (M-I) transition or crossover.  When no electrons or
{\it holes} are doped, cuprates  are insulators; they exhibit antiferromagnetism
at low temperatures. When enough electrons or holes are doped, they become metals; 
they exhibit superconductivity at low temperatures. The M-I transition or
crossover and the transition between antiferromagnetic and paramagnetic phases as
a function of doping concentrations  are asymmetric  between electron-doped and
hole-doped cuprates \cite{asymmetric}; the insulating phase is much wider in
electron-doped ones than it is in hole-doped ones, and the N\'{e}el
temperature is relatively higher in electron-doped ones than it is in
hole-doped ones. It is plausible that a relevant theory can explain not only
high-$T_{\rm c}$ superconductivity but also the asymmetry of  the M-I transition
or crossover and that of the N\'{e}el temperature.


In 1963, three distinguished theories on electron correlations in a single-band
model, which is now called the Hubbard model, were published by
Kanamori, \cite{Kanamori} Hubbard, \cite{Hubbard} and Gutzwiller. \cite{Gutzwiller}
Two theories among them are  related with the M-I transition or crossover. 
According to Hubbard's theory, \cite{Hubbard} the band splits into two subbands
called the lower and upper Hubbard bands.  According to Gutzwiller's
theory, \cite{Gutzwiller} with the help of the Fermi-liquid
theory, \cite{Luttinger1,Luttinger2} a narrow band  of quasiparticle appears on the
chemical potential;  we call them Gutzwiller's band and  Gutzwiller's
quasiparticles. The combination of the two theories implies that the density of
states must be of a three-peak structure, Gutzwiller's band  between the lower and
upper Hubbard bands.  This speculation was confirmed in a previous
paper. \cite{OhkawaSlave}  The Mott-Hubbard splitting occurs in both metallic and
insulating phases as long as the onsite repulsion $U$ is large enough, and
Gutzwiller's band is responsible for metallic behaviors.  Then, we can argue that
a metal in the vicinity of the M-I transition can become an insulator only when a
gap opens in Gutzwiller's band or that it can behave as a {\it bad}\ metal or an
insulator  when lifetime widths of Gutzwiller's quasiparticles are larger than
their bandwidth.

Not only Hubbard's \cite{Hubbard} and Gutzwiller's \cite{Gutzwiller} theories but
also the previous theory\cite{OhkawaSlave} are within the single-site approximation
(SSA). This fact implies that local electron correlations must be responsible for
the three-peak structure.   Local electron correlations are rigorously considered
in SSA that includes  all the single-site terms; such  an SSA is rigorous for 
{\it paramagnetic phases with no order parameter} in
infinite dimensions. \cite{metzner} It is reduced to solving the Anderson
model, \cite{Mapping-1,Mapping-2,georges} which is one of the simplest effective
Hamiltonians for the Kondo problem. This approximation is often called the
dynamical mean-field theory (DMFT). \cite{PhyToday} The Kondo problem has been
almost settled or almost completely
clarified \cite{singlet,poorman,Wilson,Nozieres,Yamada,Yosida}.   The Kondo
temperature $T_{\rm K}$ or $k_{\rm B}T_{\rm K}$ is defined as an energy scale of
local quantum spin fluctuations, with $k_{\rm B}$ the Boltzmann constant. 
Gutzwiller's band between the lower and upper Hubbard bands corresponds to 
the so called Kondo peak between two sub-peaks; its bandwidth
$W^*$ is about $W^*\simeq 4k_{\rm B}T_{\rm K}$.

One of the most essential physics involved in the Kondo problem is that a magnetic
moment is quenched by local quantum spin fluctuations so that the ground state is a
singlet \cite{singlet} or a normal Fermi liquid. \cite{Wilson,Nozieres}  A strongly
correlated electron liquid on a lattice is a normal Fermi liquid at $T\alt T_{\rm
K}$, where magnetic moments are quantum-mechanically quenched, while it is a
nondegenerate Fermi liquid at $T\gg T_{\rm K}$, where magnetic moments are
thermally quenched.  Local-moment magnetism occurs at $T\gg T_{\rm K}$, and
itinerant-electron one occurs at $T\alt T_{\rm K}$; superconductivity can only
occur at $T\ll T_{\rm K}$. The M-I transition or  crossover is  related with the
crossover between local-moment magnetism and itinerant-electron magnetism.

The coherence of quasiparticles is destroyed by not only thermal fluctuations but
also disorder.  When  $k_{\rm B}T\gg W^*$ or $\gamma\gg W^*$, with $\gamma$ the
quasiparticle lifetime width, quasiparticles are never well-defined; the liquid
behaves as a bad metal or an insulator. When
$k_{\rm B}T\alt W^*$ and $\gamma\alt W^*$, on the other hand, they are
well-defined; the liquid behaves as a metal.  Disorder can also play a significant
role in the M-I transition or crossover.

Kondo-lattice theories \cite{Lacroix,Zwicknagl,Coleman} have been developed
so far mainly in order to elucidate strong electron correlations in typical 
or canonical Kondo lattices of lanthanide and actinide alloys, which
are also called dense Kondo systems or heavy-fermion systems.
In these canonical Kondo lattices, almost
localized $4f$ or $5f$ electrons and itinerant conduction electrons
coexist and conduction electrons can play an important role not only 
in the quenching of magnetic moments of almost localized electrons but also 
in the appearance of an intersite exchange interaction mediated
 by conduction electrons
such as the Ruderman-Kittel-Kasuya-Yosida (RKKY) exchange interaction \cite{rkky}
and the opening of the so called hybridization gaps or pseudogaps.
In this paper, we are interested in electron correlations in
non-canonical Kondo lattices
such as transition-metal alloys in the vicinity of the M-I transition
or crossover,  where the coexistence of almost localized electrons 
and conduction electrons
is absent or it plays no essential role even if it is present.

A Kondo-lattice theory developed in this paper
is a {\it perturbation} theory starting from a starting or
an {\it unperturbed} state constructed in a {\it non-perturbative} method, 
that is,  SSA or DMFT;
intersite terms are perturbatively considered in terms of intersite
exchange interactions. This perturbation is nothing 
but $1/d$ expansion,
with $d$ being spatial dimensionality. \cite{metzner,Mapping-1,Mapping-2}
It has already been applied to typical issues on
electron correlations in non-canonical Kondo lattices
such as the Mott-Hubbard M-I transition, \cite{PhyToday}
the Curie-Weiss law of itinerant-electron magnets, \cite{ohkawaCW,miyai}
 itinerant-electron antiferromagnetism \cite{antiferromagnetism} and
ferromagnetism, \cite{ferromagnetism} 
and related issues to high-$T_ {\rm c}$ superconductivity,
\cite{OhkawaPseudogap,OhkawaEl-ph} and so on.
 Early papers \cite{Ohkawa87SC-1,Ohkawa87SC-2}  
on a mechanism of $d\gamma$-wave high-$T_{\rm c}$
superconductivity can also be regarded within the framework of
the Kondo-lattice theory.   
The interplay between correlations and disorder  in a paramagnetic phase
is investigated
for the Hubbard model with the just half filling. \cite{byczuk}
One of the purposes of this paper is to study the interplay 
between correlations and disorder
in the crossover between local-moment magnetism and itinerant-electron magnetism
as a function of electron fillings in the $t$-$J$ model. 
This paper is organized as follows:  The Kondo-lattice theory is reviewed in
\S~\ref{SecKondo-lattice}.  A paramagnetic 
unperturbed state is constructed in \S~\ref{SecFock}.
Antiferromagnetic instability of the unperturbed state
is studied in \S~\ref{SecTN}.  
Discussion is given  in \S~\ref{SecDiscussion}. Conclusion is
given in \S~\ref{SecConclusion}.  
The selfenergy of quasiparticles in disordered Kondo lattices is studied in
Appendix. 

\section{Kondo-Lattice Theory}
\label{SecKondo-lattice}

\subsection{Renormalized SSA}
\label{SecRenomalizedSSA}

We consider the $t$-$J$ model \cite{Hirsch,ZhangRice} 
or  the $t$-$t^\prime$-$J$-$U_{\infty}$  model
with $U_{\infty}/|t|\rightarrow +\infty$ on a quasi-two dimensional lattice composed of simple square lattices:
\begin{equation}\label{Eqt-Jmodel}
{\cal H} =
-t \sum_{\left< ij \right>\sigma} a_{i\sigma}^\dag a_{j\sigma}
-t^\prime \sum_{\left< ij \right>^{\prime}\sigma} 
a_{i\sigma}^\dag a_{j\sigma}
- \frac1{2} J \sum_{\left< ij \right>} 
({\bm S}_i \cdot {\bm S}_j)
+ U_{\infty} \sum_{i}n_{i\uparrow}n_{i\downarrow} , \quad 
\end{equation}
with  $t$  the transfer integral between  nearest
neighbors $\left< ij\right>$, $t^\prime$ between next-nearest
neighbors $\left< ij\right>^{\prime}$,
%
${\bm S}_i = \sum_{\alpha\beta} \frac1{2} 
\left(\sigma_x^{\alpha\beta},
\sigma_y^{\alpha\beta},
\sigma_z^{\alpha\beta} \right) 
a_{i\alpha}^\dag a_{i\beta} $,
with $\sigma_x$, $\sigma_y$, and $\sigma_z$ being the Pauli matrices,
and $n_{i\sigma}=a_{i\sigma}^\dag a_{i\sigma}$.
Infinitely large onsite repulsion $U_{\infty}$ is introduced to exclude
doubly occupied sites. The $t$-$J$ model (\ref{Eqt-Jmodel}) itself can only
treat less-than-half fillings. When the hole picture is taken, it can
also treat more-than-half fillings.  We consider two models: a {\it
symmetric} one with $t^\prime=0$, where physical properties are symmetric
between less-than-half and more-than-half fillings, and an {\it
asymmetric} one with $t^\prime\ne 0$, whose precise definition is made
in \S~\ref{SecFock}.  Effects of disorder and weak three dimensionality
are phenomenologically considered in this paper.  
 We assume $J/|t| = - 0.3 $;
$t=(0.3 \mbox{-} 0.5)$~eV and
$J =-(0.10$-$0.15)$ eV for
cuprate oxide superconductors.

We follow the previous paper \cite{OhkawaPseudogap} to treat 
$U_\infty$.   The selfenergy is divided into 
 the single-site one
$\tilde{\Sigma}_\sigma({\rm i} \varepsilon_n)$ and  the multisite
one $\Delta \Sigma_\sigma({{\rm i} \varepsilon_n,\bm k})$.   The
single-site one 
$\tilde{\Sigma}_\sigma( {\rm i} \varepsilon_n)$ is given by that of the Anderson model
with the same $U_\infty$ as that of the $t$-$J$ model (\ref{Eqt-Jmodel});
other parameters  should be  selfconsistently determined to
satisfy \cite{Mapping-1,Mapping-2} 
\begin{equation}\label{EqMap}
\tilde{G}_\sigma ({\rm i} \varepsilon_n) \!=\!
\frac1{N} \sum_{\bm k} G_\sigma({\rm i} \varepsilon_n,{\bm k}) ,
\end{equation}
with $N$ the number of sites, and  
$\tilde{G}_\sigma ({\rm i} \varepsilon_n)$ and
\begin{equation}\label{EqGreen}
G_\sigma ({\rm i} \varepsilon_{n}, {\bm k})  =
\frac1{ {\rm i}\varepsilon_n \!\!+\! \mu \!-\! E({\bm k}) 
\!-\! \tilde{\Sigma}_\sigma({\rm i}\varepsilon_n)
\!-\! \Delta\Sigma_\sigma({\rm i}\varepsilon_n,{\bm k}) }
\end{equation}
the Green functions of the Anderson and $t$-$J$  models, respectively. 
Here, $\mu$ is the chemical potential and
$E({\bm k}) = -2 t \eta_{1s}({\bm k})
-2 t^\prime \eta_{2s}({\bm k})$,
 with 
\begin{equation}
\eta_{1s}({\bm k}) = \cos(k_x a) + \cos(k_y a) ,
\end{equation}
\begin{equation}
\eta_{2s}({\bm k}) = 2\cos(k_x a) \cos(k_y a) .
\end{equation}
with $a$  the lattice constant of the simple square lattices. 

Note that the mapping condition
(\ref{EqMap}) with eq.~(\ref{EqGreen})  depends on the multisite selfenergy,
which should be perturbatively and 
selfconsistently calculated.
There can be 
various approximations of singles-site selfenergies or dynamical mean fields 
depending on it.
As is shown in \S~\ref{SecFock}, the multisite selfenergy includes 
an energy independent part, which  is denoted by
$\Delta\Sigma_\sigma({\bm k})$, because the superexchange interaction constant
does not depend on energies in the $t$-$J$ model. In this paper,
we include only 
$\Delta\Sigma_\sigma({\bm k})$ in eq.~(\ref{EqGreen});  
the Green function that includes 
$\Delta\Sigma_\sigma({\bm k})$
instead of $\Delta\Sigma_\sigma({\rm i} \varepsilon_n,{\bm k})$ 
is denoted by $G_\sigma^{(0)} ({\rm i} \varepsilon_{n}, {\bm k})$.

The Kondo temperature $T_{\rm K}$ is defined by
\begin{equation}\label{EqDefTK} 
k_{\rm B}T_{\rm K} = \left[1/\tilde{\chi}_{\rm s}(0)
\right]_{T\rightarrow 0} ,
\end{equation} 
where $\tilde{\chi}_{\rm s}({\rm i} \omega_l)$ is
the spin susceptibility of the Anderson model,
which does not include the conventional factor
$(1/4)g^2\mu_{\rm B}^2$, 
with $g$ and $\mu_{\rm B}$ being the $g$ factor and
the Bohr magneton, respectively.
The single-site selfenergy 
$\tilde{\Sigma}_\sigma({\rm i} \varepsilon_n)$ is expanded  for 
$|\varepsilon_n|\alt 2k_{\rm B}T_{\rm K}$ as
$\tilde{\Sigma}_\sigma({\rm i} \varepsilon_n) =
\tilde{\Sigma}(0) + (1 - \tilde{\phi}_\gamma) {\rm i} \varepsilon_n
+ \sum_{\sigma^\prime}(1 -
\tilde{\phi}_{\sigma\sigma^\prime}) 
\Delta\mu_{\sigma^\prime} + \cdots$,
with 
$\tilde{\phi}_\gamma=\tilde{\phi}_{\sigma\sigma}$
and $\Delta\mu_{\sigma}$ infinitesimally small spin-dependent chemical potential
shifts. The Wilson ratio is defined by
$\tilde{W}_{\rm s} = \tilde{\phi}_{\rm s}/\tilde{\phi}_\gamma$,
with 
$\tilde{\phi}_{\rm s}=
\tilde{\phi}_{\sigma\sigma}-\tilde{\phi}_{\sigma-\sigma}$.
For almost half fillings, charge fluctuations are suppressed so that
$\tilde{\phi}_{\rm c}=
\tilde{\phi}_{\sigma\sigma}+\tilde{\phi}_{\sigma-\sigma} \ll 1$.
For such fillings, 
$\tilde{\phi}_\gamma \gg 1$
so that $\tilde{W}_{\rm s} \simeq 2$. 
The Green function is divided into coherent and
incoherent parts:
%
$G_\sigma^{(0)} ({\rm i} \varepsilon_{n}, {\bm k}) =
(1/\tilde{\phi}_\gamma)
g_\sigma^{(0)} ({\rm i} \varepsilon_{n}, {\bm k})
\!+\! \mbox{(incoherent part)} $, 
%
with
\begin{equation}\label{EqGreenCoh}
g_\sigma^{(0)} ({\rm i} \varepsilon_{n}, {\bm k}) =
\frac1{{\rm i} \varepsilon_{n} +\mu^* 
- \xi ({\bm k}) + {\rm i} \gamma 
\varepsilon_{n}/|\varepsilon_{n}|} ,
\end{equation}
with $\mu^* =(\mu-\tilde{\Sigma}_0)/\tilde{\phi}_\gamma$ and 
$\xi ({\bm k}) =[E({\bm k})+\Delta\Sigma({\bm k})]/\tilde{\phi}_\gamma$.
 We introduce a phenomenological lifetime width $\gamma$. Although $\gamma$
depends on energies even if it is due to disorder, as is discussed in
Appendix~\ref{SecDisorder}, its energy dependence is ignored. \cite{ComGamma}

According to the Fermi-surface sum rule, \cite{Luttinger1,Luttinger2}
the number of electrons per site  for $T/T_{\rm K} \rightarrow
+0$ and 
$\gamma/k_{\rm B}T_{\rm K} \rightarrow +0$ is given by
\begin{equation}\label{EqFSumRule}
n =
2 \int \! d\varepsilon \rho_{\gamma\rightarrow 0}(\varepsilon) 
f_\gamma(\varepsilon - \mu^*) 
%
= 2 \int \! d\varepsilon \rho_{\gamma}(\varepsilon) 
f_{\gamma=0}(\varepsilon - \mu^*) ,
\end{equation}
with
\begin{equation}
\rho_\gamma (\varepsilon) = \frac1{\pi N} \sum_{\bm k}
\frac{\gamma}
{\left[\varepsilon - \xi({\bm k})\right]^2 +\gamma^2} , \qquad 
\end{equation}
\begin{equation}\label{EqfGamma}
f_\gamma(\varepsilon)  =
\frac1{2} +\frac1{\pi} \mbox{Im}  \left[
\psi\left( \frac1{2} 
+ \frac{\gamma - {\rm i} \varepsilon}{2\pi k_{\rm B} T} \right)
\right] ,
\end{equation}
with $\psi (z)$ the di-gamma function. Note that
$f_{\gamma=0}(\varepsilon) =1/\left[\exp(\varepsilon/k_{\rm B}T)+1\right] $.
We assume eq.~(\ref{EqFSumRule}) even for nonzero 
$T$ and $\gamma$. The parameter $\tilde{\Sigma}_0$
or $\mu^*$ can be determined from  eq.~(\ref{EqFSumRule}) as a function
of $n$.

\subsection{Intersite exchange interaction}

In Kondo lattices, local spin fluctuations at different
sites interact with each other by an exchange interaction.  Following
this physical picture, we define an intersite exchange interaction
$I_{\rm s}({\rm i} \omega_l, {\bm q})$ by 
\begin{equation}\label{EqKondoSus}
\chi_{\rm s}({\rm i} \omega_l, {\bm q}) =
\frac{\tilde{\chi}_{\rm s}({\rm i} \omega_l)}
{1 - \mbox{$\frac{1}{4}$}I_{\rm s}({\rm i} \omega_l, {\bm q})
\tilde{\chi}_{\rm s}({\rm i}\omega_l)} ,
\end{equation}
where  $\chi_{\rm s}({\rm i} \omega_l, {\bm q})$ is
the spin susceptibility of the $t$-$J$ model to be studied.
Following the previous paper, \cite{OhkawaPseudogap} we obtain 
\begin{equation}\label{EqIs}
I_{\rm s} ({\rm i} \omega_l, {\bm q}) = J({\bm q}) + 
2 U_\infty^2 \Delta\pi_{\rm s}({\rm i} \omega_l, {\bm q}) ,
\end{equation}
with $J({\bm q}) =2J\eta_{1s}({\bm q})$  and 
$\Delta\pi_{\rm s}({\rm i} \omega_l, {\bm q})$ the multi-site part of the
irreducible polarization function in spin channels; 
$2 U_\infty^2 \Delta\pi_{\rm s}({\rm i} \omega_l, {\bm q})$ is examined
in \S~\ref{SecExc} and \S~\ref{SecMode}.

According to the Ward relation, \cite{ward} 
the irreducible single-site three-point vertex function in spin channels, 
$\tilde{\lambda}_{\rm s}({\rm i} \varepsilon_n,{\rm i} \varepsilon_n \!+\! {\rm i} \omega_l;{\rm i} \omega_l)$,
is given by
\begin{equation}\label{EqThreeL}
 U_\infty 
\tilde{\lambda}_{\rm s} ({\rm i} \varepsilon_n,{\rm i} \varepsilon_n+{\rm i} \omega_l;{\rm i} \omega_l)
= 2\tilde{\phi}_{\rm s} / \tilde{\chi}_{\rm s}({\rm i} \omega_l)  ,
\end{equation}
for $|\varepsilon_n| \rightarrow +0$ and 
$|\omega_l| \rightarrow +0$.
 We approximately use eq.~(\ref{EqThreeL}) for 
$|\varepsilon_n| \alt 2 k_{\rm B}T_{\rm K}$ and
$|\omega_l| \alt 2 k_{\rm B}T_{\rm K}$.
A mutual interaction between quasiparticles mediated by spin fluctuations is
given by
\begin{equation}\label{SpinFluctI}
\frac{1}{4} \!\left[
2\tilde{\phi}_{\rm s} / \tilde{\chi}_{\rm s}({\rm i} \omega_l)\right]^2 
 F_{\rm s}({\rm i} \omega_l,{\bm q}) = 
\tilde{\phi}_{\rm s}^2 \frac{1}{4} I_{\rm s}^* ({\rm i} \omega_l, {\bm q}) ,
\end{equation}
with 
\begin{equation}\label{EqF}
 F_{\rm s}({\rm i} \omega_l,{\bm q}) =
\chi_{\rm s}({\rm i} \omega_l,{\bm q}) - \tilde{\chi}_{\rm s}({\rm i} \omega_l) ,
\end{equation}
\begin{equation}\label{EqIs*2}
\frac{1}{4} I_{\rm s}^*({\rm i} \omega_l, {\bm q}) =
\frac{ \frac{1}{4}
I_{\rm s} ({\rm i} \omega_l, {\bm q}) }
{1 - \frac{1}{4}I_{\rm s}({\rm i} \omega_l, {\bm q})
\tilde{\chi}_{\rm s}({\rm i} \omega_l) } .
\end{equation}
The single-site term is subtracted in $ F_{\rm s}({\rm i} \omega_l,{\bm q})$ because it
is considered in SSA. The expansion coefficient $\tilde{\phi}_{\rm s}$
appears as an effective or reducible single-site
vertex function in eq.~(\ref{SpinFluctI}).
Because the spin space is isotropic,  the interaction in the transversal channels
is also given by these equations. 
 In the 
Kondo-lattice theory, intersite effects are perturbatively considered in terms of
 $I_{\rm s}({\rm i} \omega_l, {\bm q})$,  $I_{\rm s}^*({\rm i} \omega_l, {\bm q})$, or 
  $F_{\rm s}({\rm i} \omega_l,{\bm q})$
depending on each situation.

\section{Renormalization of Gutzwiller's Quasiparticles}
\label{SecFock}
In this section, we  restrict the Hilbert space within paramagnetic states. 
There are two types of selfenergies linear in the exchange interaction: 
Hartree-type and Fock-type terms.  The Hartree-type term vanishes in paramagnetic
states. We consider the Fock-type term of
the superexchange interaction: 
\begin{equation}\label{EqSelf-1}
\Delta \Sigma_\sigma ({\bm k}) \!=\!
\frac{3}{4} \tilde{\phi}_{\rm s}^2
\frac{k_{\rm B} T}{N} \!\!
\sum_{\varepsilon_{n}{\bm p}}
\! J({\bm k}\!-\!{\bm p}) 
e^{{\rm i} \varepsilon_{n}0^+}\!
G_\sigma ({\rm i} \varepsilon_{n}, {\bm p}) .
\end{equation}
The factor 3 appears because of three
spin channels. 
When only the coherent part is considered,  
\begin{equation}
\frac1{\tilde{\phi}_\gamma}\Delta \Sigma_\sigma ({\bm k}) =
\frac{3}{4} \tilde{W}_{\rm s}^2
J \Xi \eta_{1s}({\bm k})  , 
\end{equation}
with
\begin{equation}\label{EqXi}
\Xi =
\frac1{ N}\sum_{\bm k} \eta_{s} ({\bm k}) 
f_\gamma\left[\xi({\bm k})-\mu^* \right] .
\end{equation}
The dispersion relation of quasiparticles is given
by   
\begin{equation}
\xi({\bm k}) = 
- 2t^* \eta_{1s}({\bm k}) 
- 2t_2^* \eta_{2s}({\bm k}) ,
\end{equation} 
where $t^*$  should be
selfconsistently determined to satisfy
\begin{equation}\label{Eqt*}
2 t^* = \frac{2t}{\tilde{\phi}_\gamma} 
- \frac{3}{4}\tilde{W}_{\rm s}^2 J \Xi, 
\end{equation}
and $t_2^*$ is simply given by
$t_2^* = t^\prime/\tilde{\phi}_\gamma $.

For the symmetric model, $t^\prime=0$ so that $t_2^*=0$. 
In order to examine how crucial role the shape of the Fermi
surface plays in the asymmetry, we consider a phenomenological
asymmetric model with
$t_2^*/t^* = - 0.3 $. 

\begin{figure*}
\centerline{\hspace*{0.8cm}
\includegraphics[width=4.8cm]{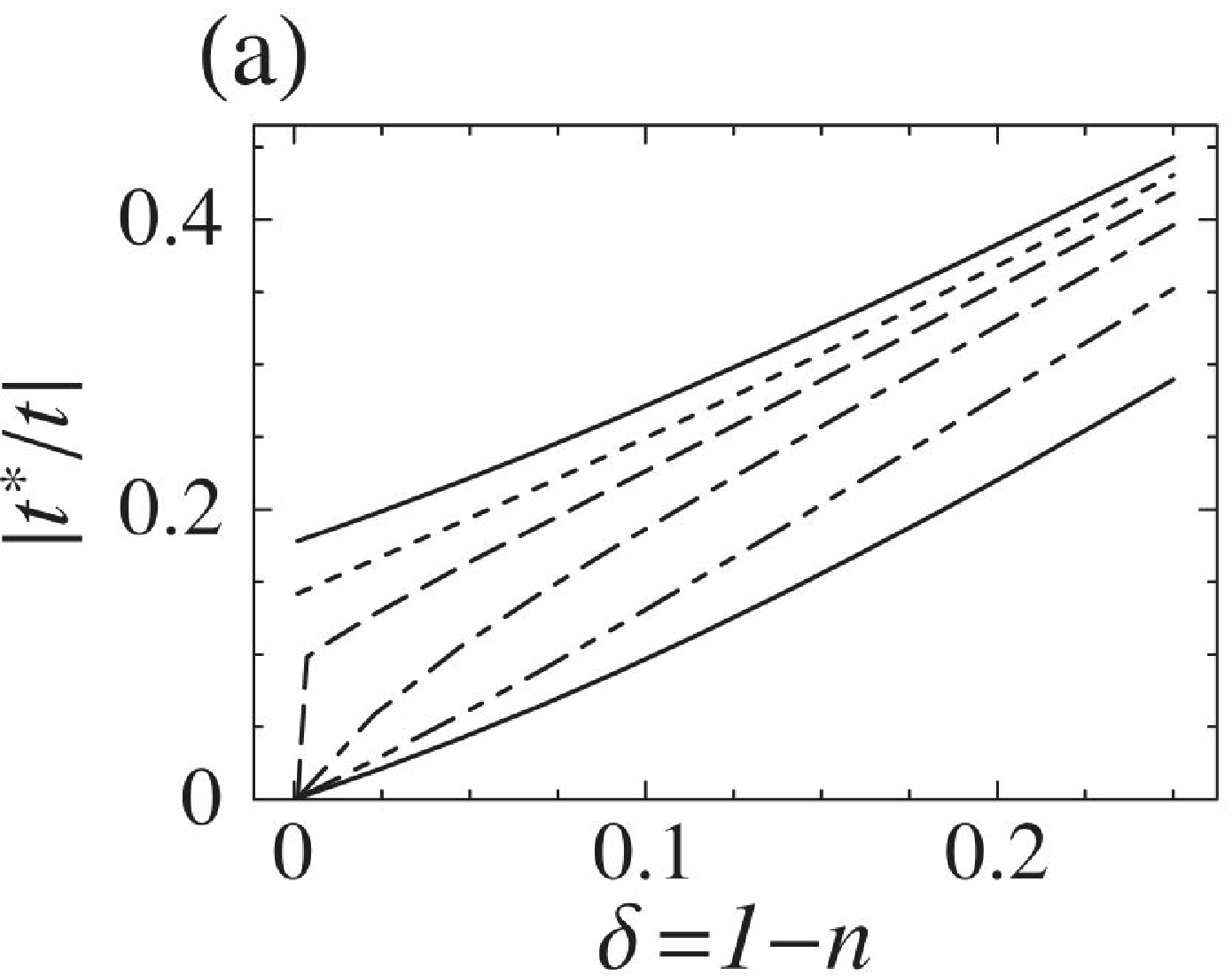}\hspace{-0.3cm}%
\includegraphics[width=4.8cm]{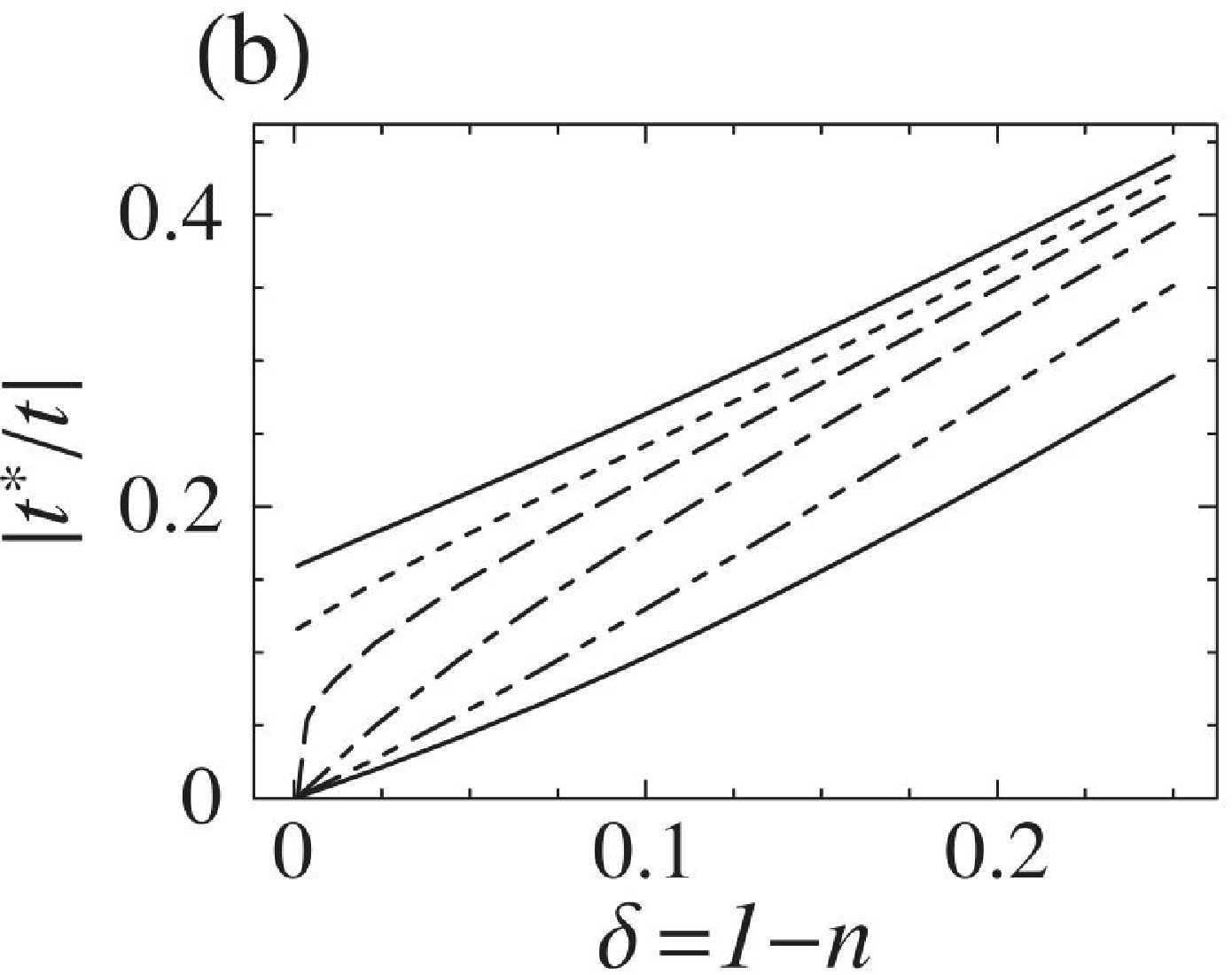}\hspace{-0.3cm}%
\includegraphics[width=4.8cm]{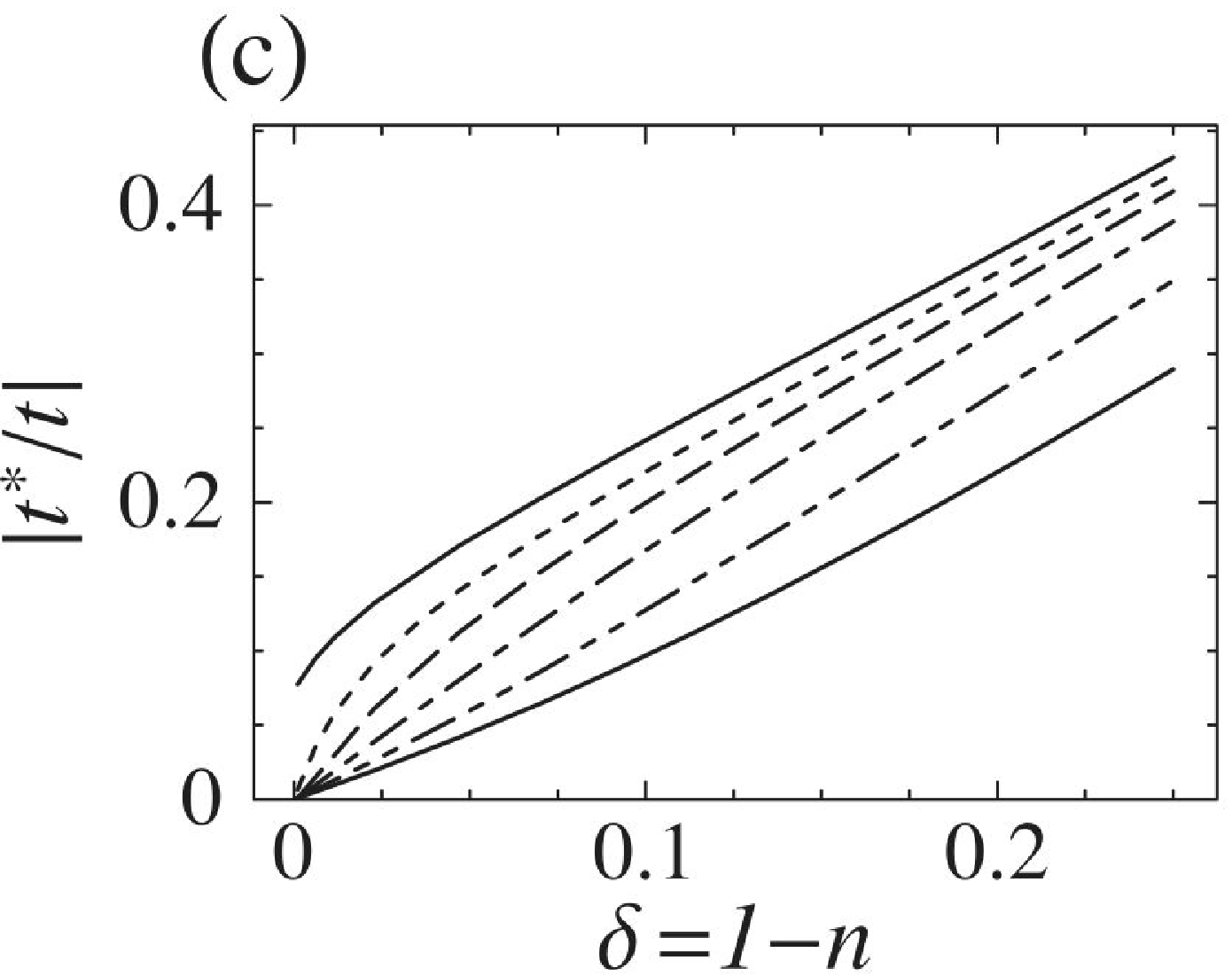}\hspace{-0.3cm}%
\includegraphics[width=4.8cm]{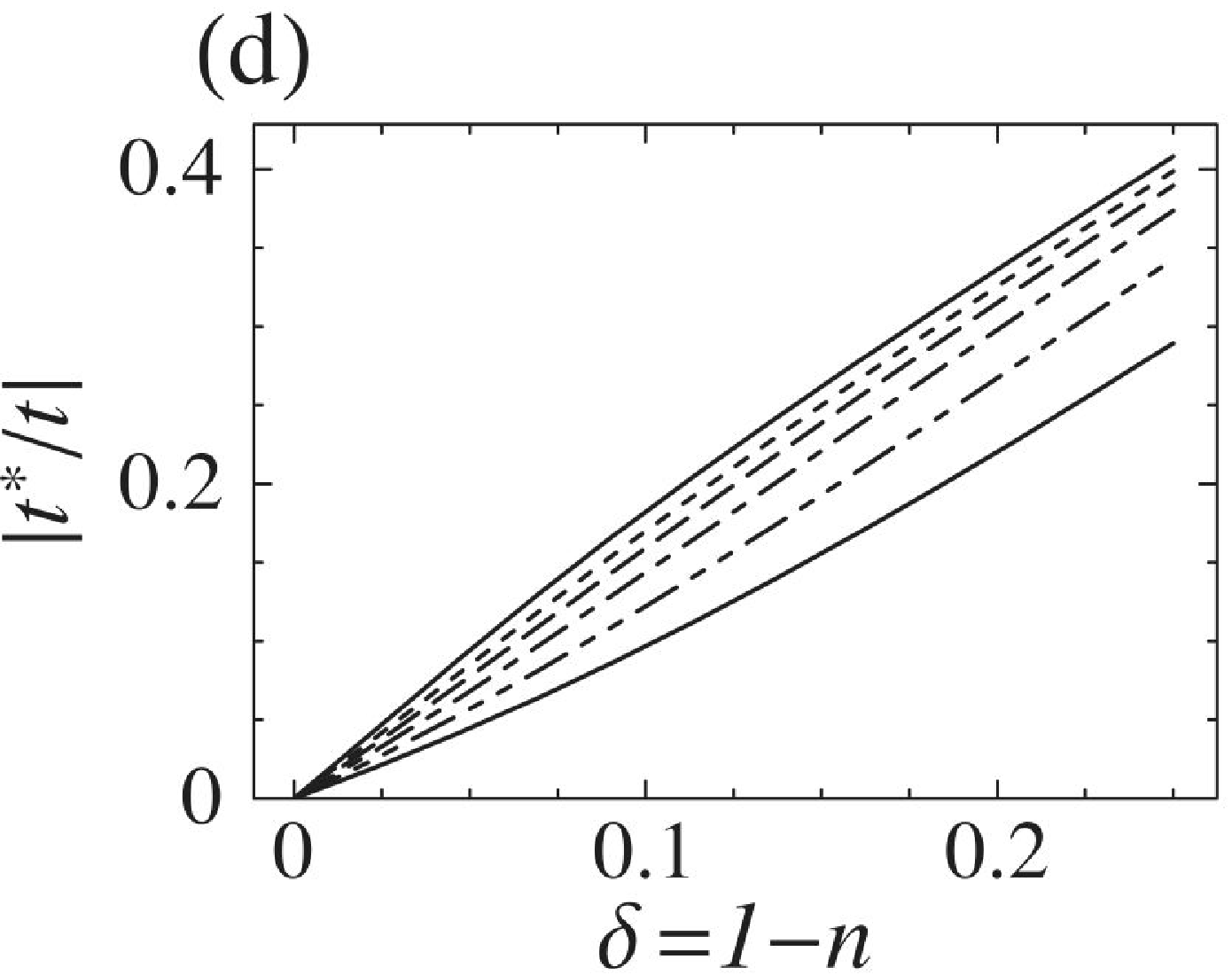}
}
\caption[1]{
Renormalized transfer integrals $t^*$ of the symmetric
model: (a) $k_{\rm B}T/|t|=0.02$, (b) $k_{\rm B}T/|t|=0.1$,
(c) $k_{\rm B}T/|t|=0.2$, and
(d) $k_{\rm B}T/|t|=0.4$.
In each figure, topmost solid, dashed, broken,
chain, and chain double-dashed lines show results for
$\gamma/|t|=0.01$,  0.1, 0.2,  0.4, and 1, respectively.  
For comparison,
$1/\tilde{\phi}_\gamma$ is also shown by a bottom solid line.
}
\label{t-star}
\end{figure*}

\begin{figure}
\centerline{
\includegraphics[width=8cm]{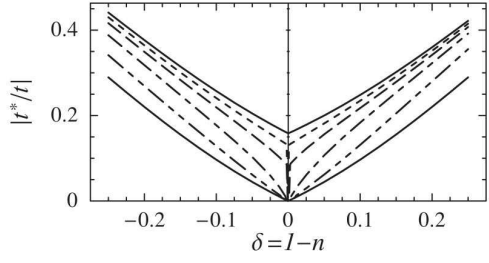}
}
\caption[2]{
$t^*$ of the asymmetric model for $k_{\rm B}T/|t|=0.02$.  See also the
caption of Fig.~\ref{t-star}.
 }
\label{t-star-asym}
\end{figure}

\begin{figure*}
\centerline{\hspace*{0.5cm}
\includegraphics[width=5.6cm]{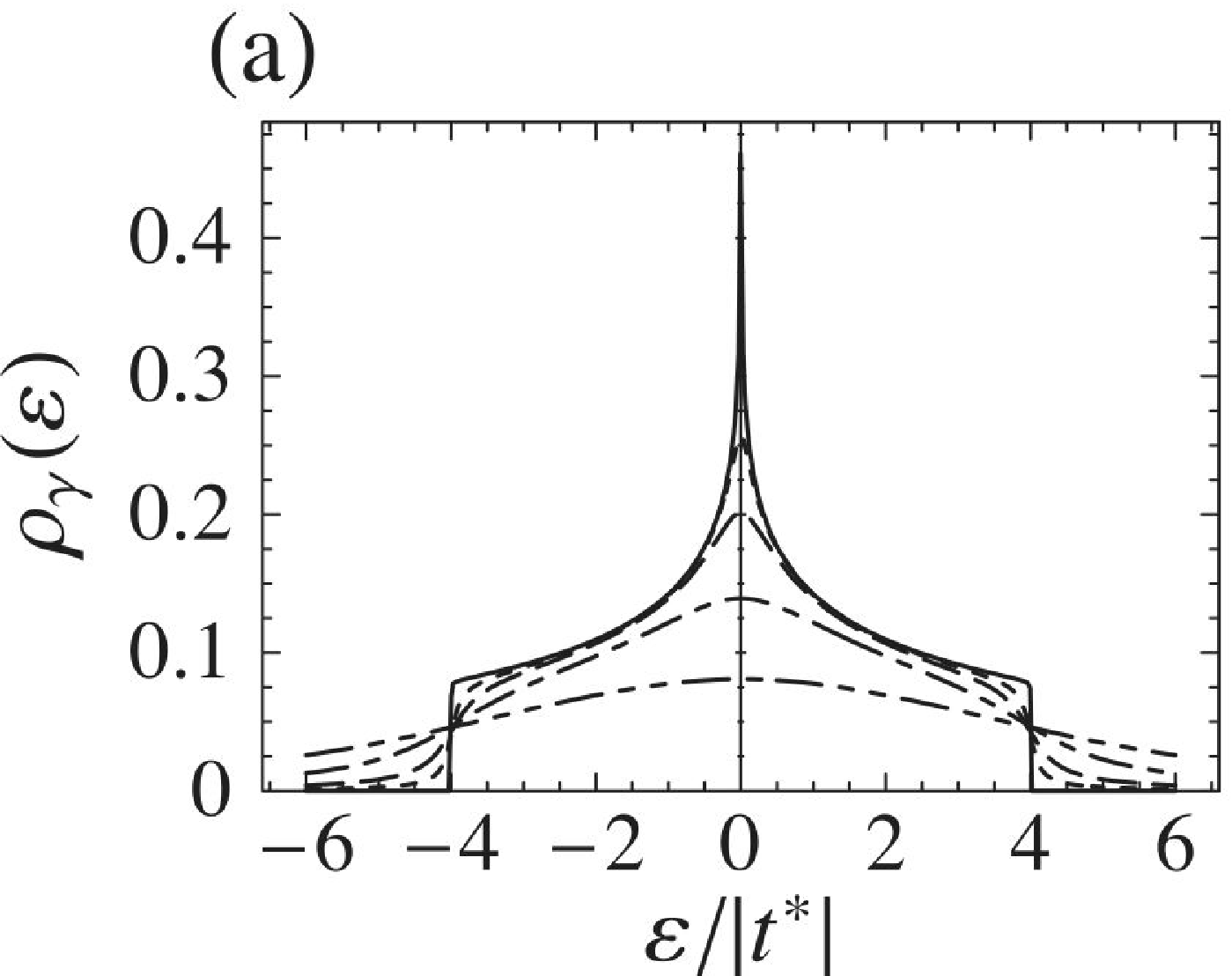}\hspace{-0.9cm}%
\includegraphics[width=5.0cm]{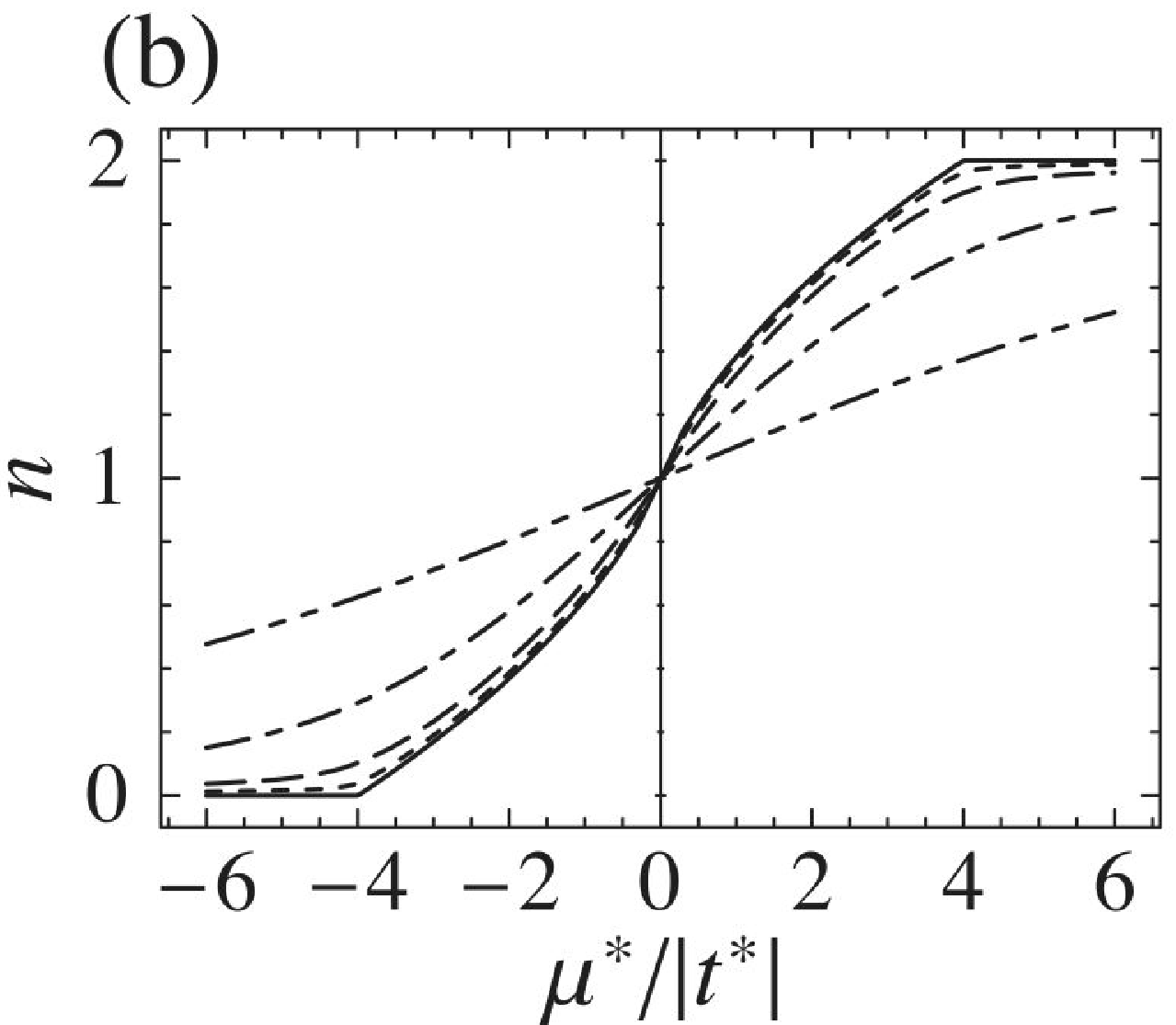} \hspace{-0.6cm}%
\includegraphics[width=5.1cm]{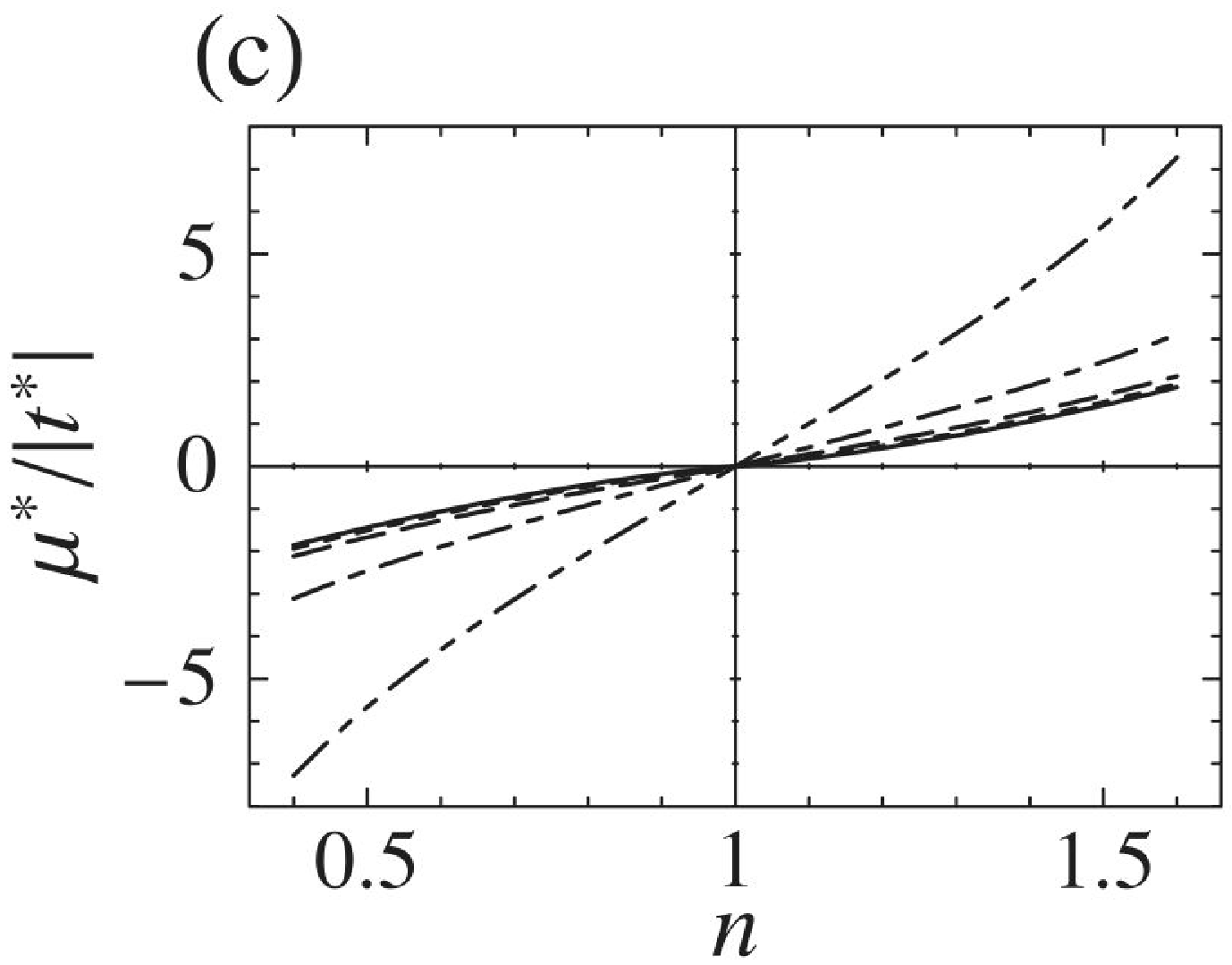} \hspace{-0.5cm}%
\includegraphics[width=4.3cm]{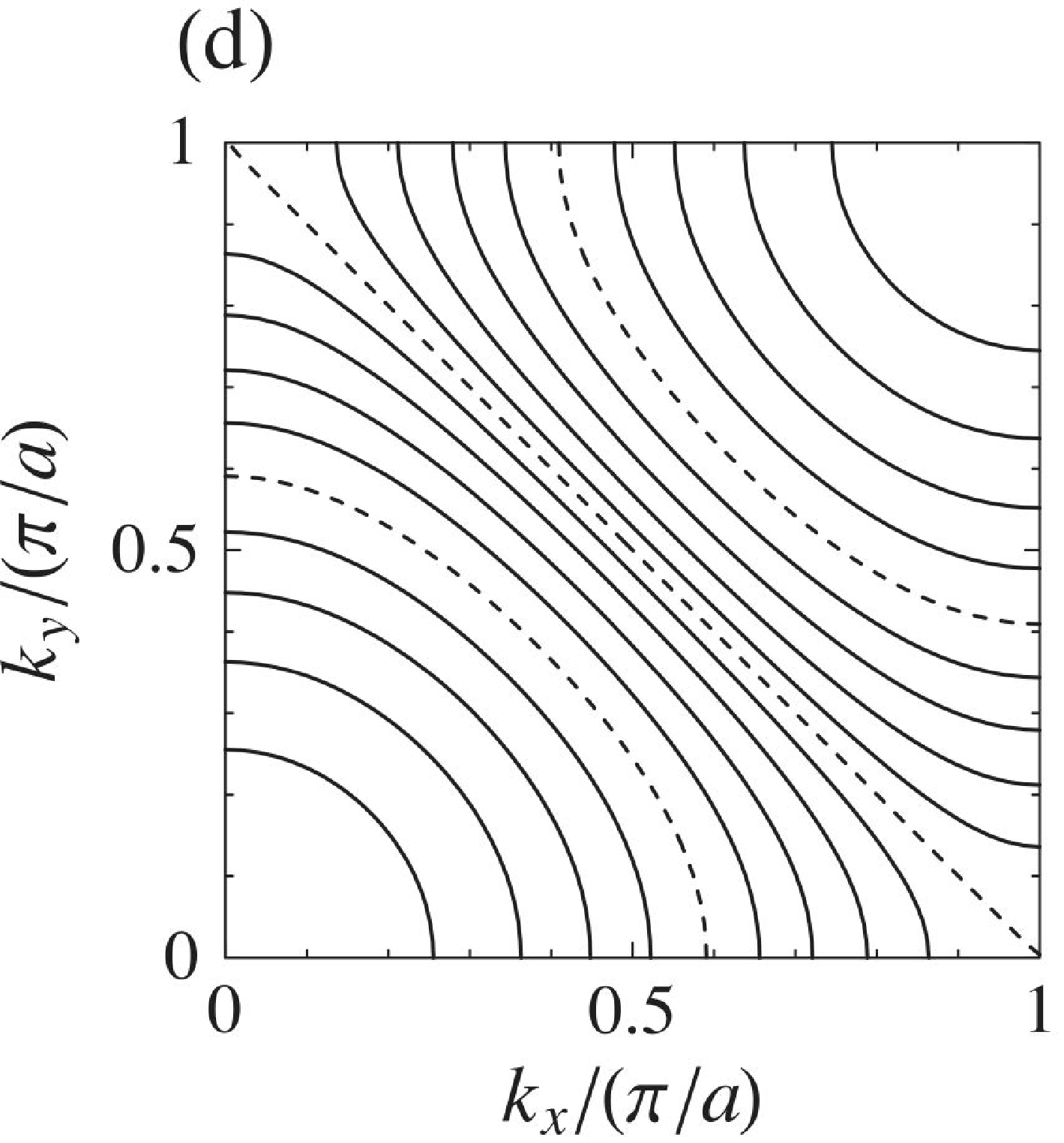}
}
\caption[3]{
Single-particle properties of the unperturbed state of the
symmetric model.   (a)
Density of states for quasiparticles
$\rho_\gamma(\varepsilon)$,  
(b) effective chemical potentials, $\mu^*$, as functions of
carrier concentrations $n$, and
(c) $n$ as functions of $\mu^*$. 
In these three figures, solid, dashed, broken, chain, and
chain double-dashed lines show results for
$\gamma/|t^*|=10^{-3}$,  0.1, 0.3, 1, and 3, respectively.
 Note that the solid and dashed lines are the
almost same as each other. (d) Fermi surfaces for $k_{\rm B}T/|t|=0$,
$\gamma=0$, and 19 electron concentrations such as  $n=0.1 \times
i$, with $1\le i \le 19$ being an integer. Dashed lines show
Fermi surfaces for  $n=$0.5, 1.0, and 1.5. Single-particle
properties of the asymmetric model can be found in Fig.~2 of
ref.~\citeonline{OhkawaPseudogap}.
 }
\label{rho_n_mu_FS}
\end{figure*}

Expansion parameters $\tilde{\phi}_\gamma$ and $\tilde{\phi}_{\rm s}$ 
are given by those of  the mapped Anderson model.  However, we
approximately use those for the Anderson model with a constant
hybridization energy. According to Appendix of the previous
paper \cite{ferromagnetism}, 
\begin{equation}\label{EqPhiG}
\tilde{\phi}_\gamma= \frac1{2}
\left(\frac{1}{|\delta|} + |\delta|\right)
\frac{\displaystyle \left(\pi /2 \right)^2(1-|\delta|)^2}
{\displaystyle \cos^2 \!\left(\pi \delta/2 \right)} ,
\end{equation}
\begin{equation}\label{EqPhiS}
\tilde{\phi}_{\rm s} = \frac{1}{|\delta|} 
\frac{\displaystyle \left(\pi /2 \right)^2(1-|\delta|)^2}
{\displaystyle \cos^2 \!\left(\pi \delta/2 \right)} ,
\end{equation}
where 
$\delta = 1 - n$
is the concentration of dopants, {\it holes}  ($\delta>0$) or
electrons ($\delta<0$). These are consistent with 
Gutzwiller's theory. \cite{Gutzwiller}
Figures~\ref{t-star} and \ref{t-star-asym} show $t^*$ of the symmetric and
asymmetric models as a function of $\delta$ for various $\gamma$.  It is interesting that $t^*$ is nonzero even
for $\delta\rightarrow 0$ if $\gamma$ and $k_{\rm B}T$ are small enough.  For
the symmetric model ($t_2^* = 0$),  eq.~(\ref{EqXi}) can be analytically
calculated for $T =0$~K, $\gamma =0$ and  $\delta =0$ $(\mu^* =0)$ in such a way
that $\Xi=4/\pi^2 $. It follows that 
$\left[t^* / t \right]_{\delta \rightarrow 0}
\rightarrow -\left( 3\tilde{W}_{\rm s}^2J / 2\pi^2  t \right) =  0.18 $
for $J/t=-0.3$. If $\gamma$ or
$k_{\rm B}T$ is large enough, on the other hand, $\Xi$ and $t^*$ vanish for
$\delta\rightarrow 0$.

Figure~\ref{rho_n_mu_FS} shows physical properties of the unperturbed
state of the symmetric model. Those of the unperturbed
state of the asymmetric model can be found in Fig.~2 of
ref.~\citeonline{OhkawaPseudogap}.
In \S~\ref{SecTN}, we 
study antiferromagnetic instabilities of these paramagnetic unperturbed states.


\section{Antiferromagnetic Instability} 
\label{SecTN}
\subsection{Reduction of $T_{\rm N}$ by local spin fluctuations}
\label{SecTN1}
 According to eq.~(\ref{EqKondoSus}), 
the N\'{e}el temperature $T_{\rm N}$ is determined by 
the competition between the Kondo effect and the intersite exchange
interaction:
\begin{equation}\label{EqAFCondition}
\frac1{\tilde{\chi}_{\rm s}(0)} - \frac1{4} I_{\rm s}(0, {\bm Q}) = 0,
\end{equation}
where ${\bm Q}$ is an ordering wave number to be determined. 
Effects of local spin fluctuations in SSA or those of dynamical mean fields 
in DMFT are
included in the local term of $1/\tilde{\chi}_{\rm s}(0)$;  
$I_{\rm s}(0, {\bm Q})$  corresponds to the conventional
Weiss's mean field, 
 which is an intersite effect and can only be included beyond SSA.

The local susceptibility $\tilde{\chi}_{\rm s}(0)$ is almost constant at 
$T \ll T_{\rm K}$, while it obeys the Curie law at $T\gg T_{\rm K}$. 
We use an interpolation between the two limits:
\begin{equation}\label{EqCW-local}
\tilde{\chi}_{\rm s}(0) = \frac{n}
{\displaystyle k_{\rm B} \sqrt{n^2T_{\rm K}^2+ T^2}} .
\end{equation}
Here, $T_{\rm K}$ is an averaged one over  $T_{\rm K}$'s at different sites.
In disordered systems, the
mapping conditions are different from site to site so that $T_{\rm K}$'s are also
different from site to site. Such disorder in $T_{\rm K}$ causes 
energy-dependent lifetime widths, as is studied in
Appendix. However, lifetime widths due to the
disorder in $T_{\rm K}$ are small on the chemical potential in case of
non-magnetic impurities. 
According to eqs.~(\ref{EqMap}) and (\ref{EqDefTK})
together with the Fermi-liquid  relation, \cite{Yamada,Yosida} it follows that 
%
%
\begin{equation}\label{EqTK2}
1/k_{\rm B}T_{\rm K} 
= 2 \tilde{W}_{\rm s} \left[\rho_{\gamma} (\mu^*)
\right]_{\gamma \rightarrow 0}, 
\end{equation}
in the absence of disorder.  
Then, a mean value of $T_{\rm K}$ in disordered systems
is approximately given by eq.~(\ref{EqTK2}) with nonzero but
small $\gamma$.  It follows from eq.~(\ref{EqTK2}) that
\begin{equation}\label{EqTKdef}
k_{\rm B}T_{\rm K} = \frac{|t^*|}{c_{T_{\rm K}}}\frac{ 1}{2\tilde{W}_{\rm s}} ,
\end{equation}
with $c_{T_{\rm K}}$  a numerical constant depending on $n$. 
As is shown in Fig.~\ref{rho_n_mu_FS}(a), 
$\rho_\gamma(\mu^*) \simeq 0.15$ for 
$0.1 \alt \gamma/|t^*| \alt 1$ and
$0.05 \alt |\delta| \alt 0.25$.
We assume that  $c_{T_{\rm K}}$ is independent of $n$ for the sake of
simplicity:  
$c_{T_{\rm K}}=0.15 $.
We are only interested in physical properties that never drastically
change when  $c_{T_{\rm K}}$ slightly changes.

\subsection{Exchange interaction arising from the exchange of
pair excitations of quasiparticles}
\label{SecExc}

The second term of eq.~(\ref{EqIs}) is the sum of  an exchange
interaction arising from the virtual exchange of pair excitations of
quasiparticles, 
$J_{\rm Q}({\rm i} \omega_l, {\bm q})$, and  the mode-mode coupling term,
$-4\Lambda ({\rm i} \omega_l, {\bm q})$:
\begin{equation}\label{EqJQ-Lamda}
2 U_\infty^2 \Delta\pi_{\rm s}({\rm i} \omega_l, {\bm q}) =
J_{\rm Q}({\rm i} \omega_l, {\bm q}) - 4 \Lambda ({\rm i} \omega_l, {\bm q}).
\end{equation}
 When higher-order terms in intersite effects are ignored,  
\begin{equation}\label{EqJQ1}
J_{\rm Q}({\rm i} \omega_l, {\bm q}) =
4\left[\frac{\tilde{W}_{\rm s}}{\tilde{\chi}_{\rm s}(0) }\right]^2 
\left[ P({\rm i} \omega_l, {\bm q}) - P_0({\rm i} \omega_l)
\right], 
\end{equation}
with
\begin{equation}\label{EqP}
P({\rm i} \omega_l, {\bm q}) =
\frac{k_{\rm B}T}{N} \!\!\sum_{\varepsilon_n{\bm k}\sigma} \!
g_\sigma^{(0)}({\rm i} \varepsilon_n \!+\! {\rm i} \omega_l, {\bm k}
 \!+\! {\bm q})  g_\sigma^{(0)}({\rm i} \varepsilon_n, {\bm k}) .
\end{equation}
The local contribution 
$P_0({\rm i} \omega_l) =(1/N)\sum_{\bm q} P({\rm i} \varepsilon_n, {\bm q})$ is
subtracted because it is considered in SSA. 
The static component is
simply given by
\begin{equation}
P(0,{\bm q}) =
\frac{2}{N} \!\sum_{\bm k}\! \frac{f_\gamma 
\bigl[\xi({\bm k} \!+\! {\bm q}) \!-\! \mu^*\bigr]
\!-\! f_\gamma \bigl[\xi({\bm k}) \!-\! \mu^*\bigr]}
{ \xi({\bm k}) -\xi({\bm k} +{\bm q}) -{\rm i} 0} .
\end{equation}
The magnitude of $J_{\rm Q}(0, {\bm q})$ is proportional to $k_{\rm B}T_{\rm K}$
or the quasiparticle bandwidth.

\begin{figure*}
\centerline{\hspace*{0.7cm}
\includegraphics[width=5.8cm]{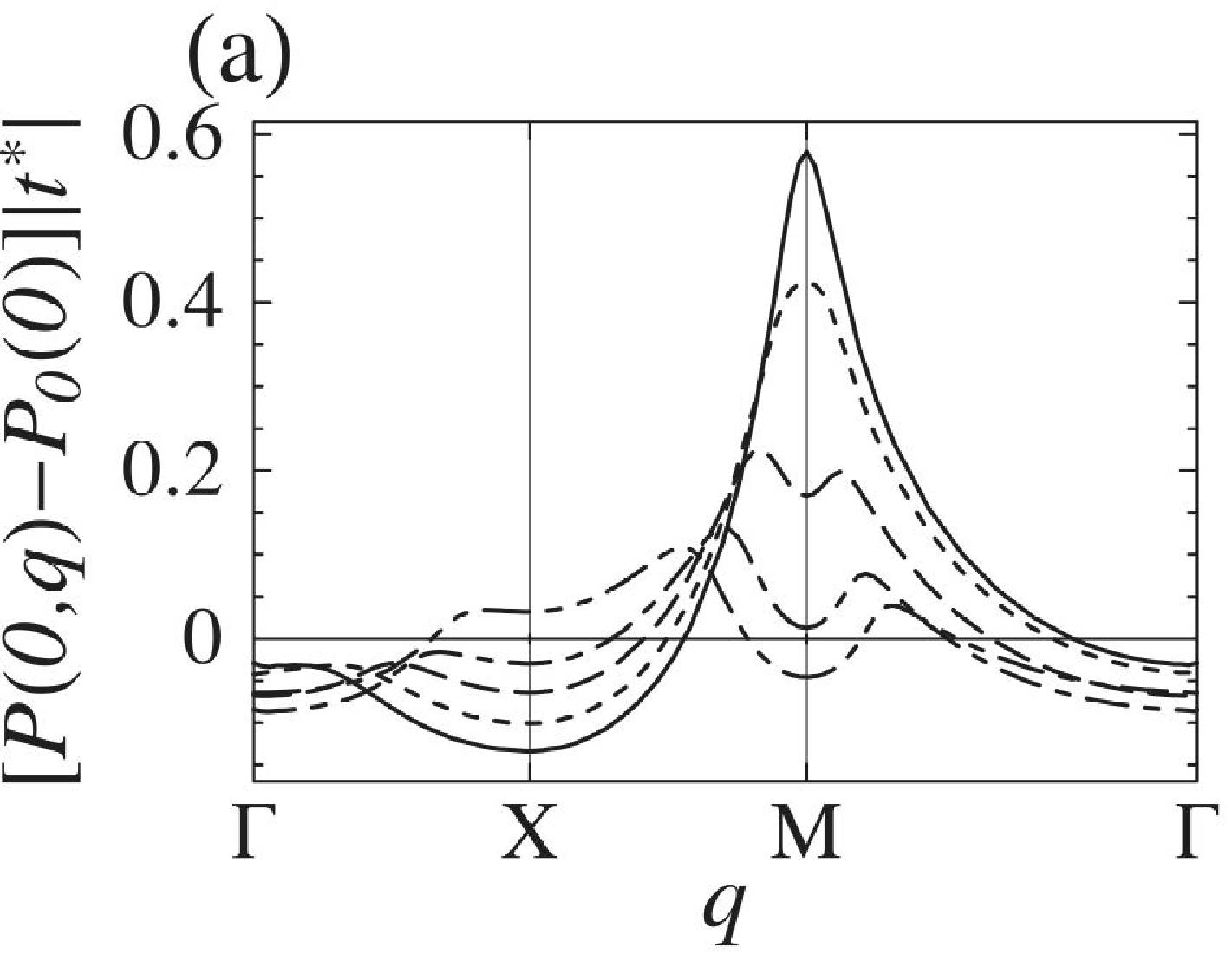}
\includegraphics[width=5.8cm]{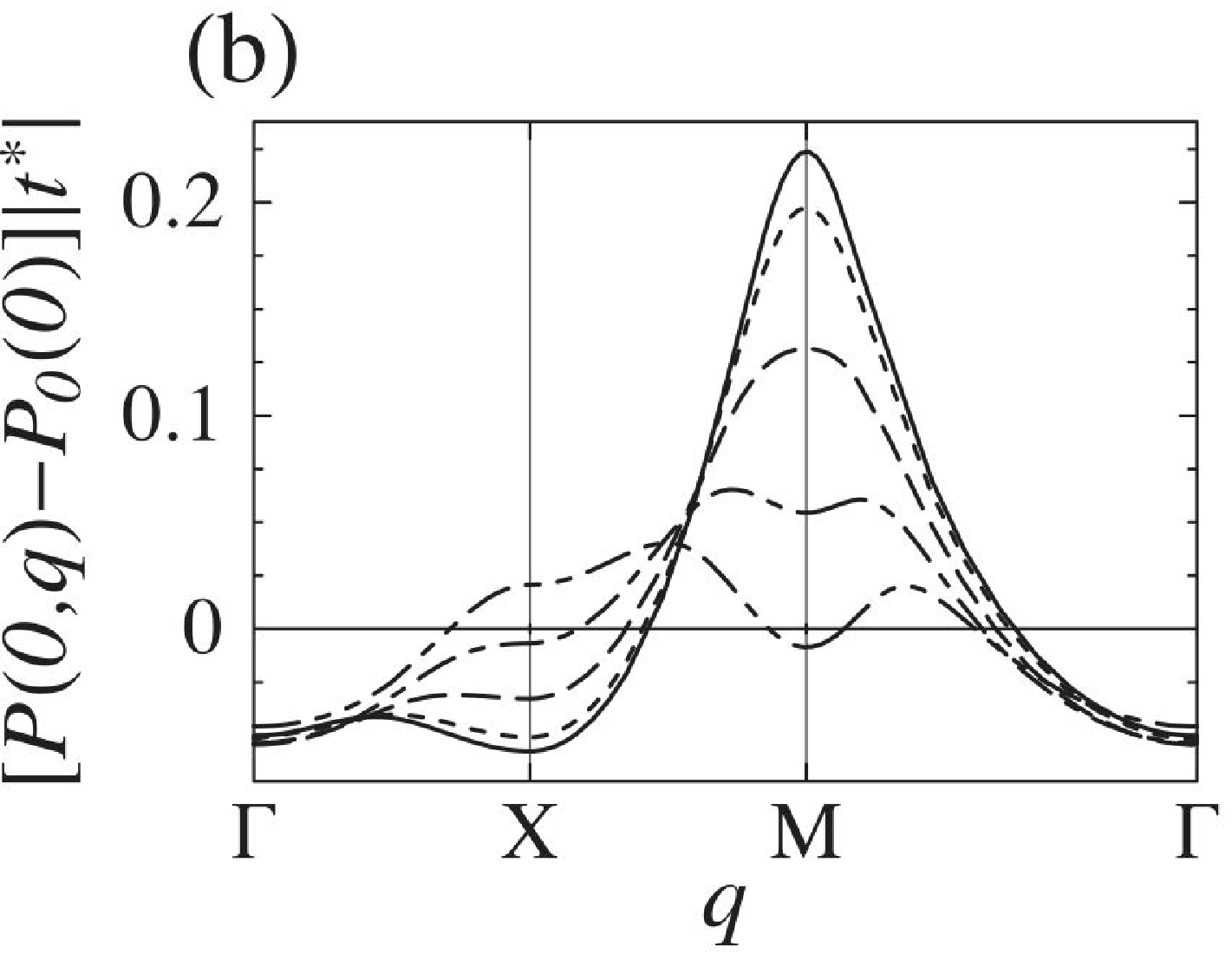}
\includegraphics[width=6.2cm]{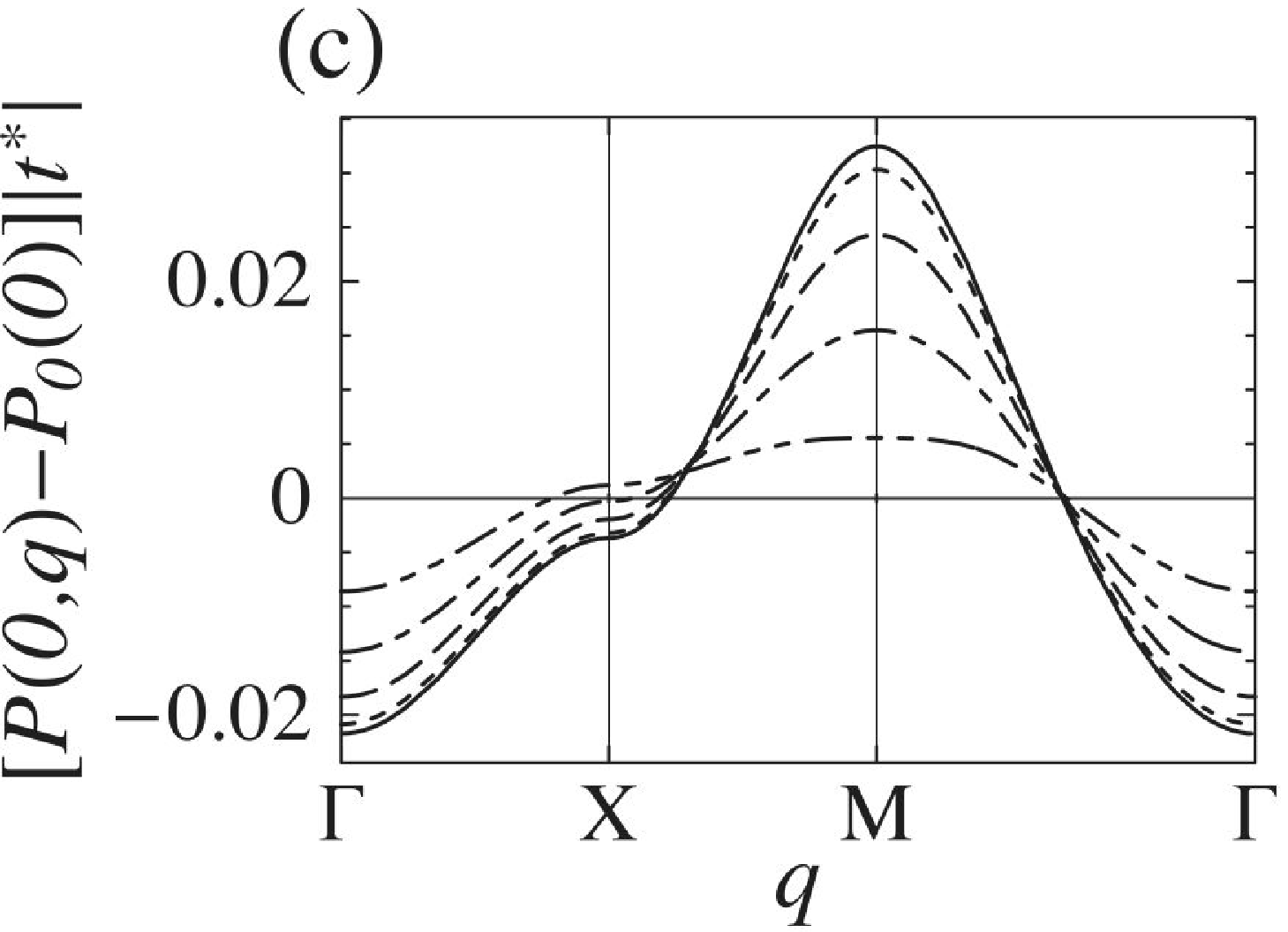}
}
\caption[4]{ 
Static polarization function
$[P(0,{\bm q})-P_0(0)]|t^*|$ of the symmetric model:
(a) $k_{\rm B}T/|t^*|=\gamma/|t^*|=0.1$,
(b) $k_{\rm B}T/|t^*|=\gamma/|t^*|=0.3$, and
(c) $k_{\rm B}T/|t^*|=\gamma/|t^*|=1$.
Solid, dashed, broken, chain, and chain double-dashed lines show results
for $n \rightarrow 1$, $n=0.9$, 0.8, 0.7, and 0.6, respectively.
Here, $\Gamma$, $X$ and $M$ stand for $(0,0)$, $(\pi/a,0)$ and
$(\pi/a,\pi/a)$, respectively.
}
\label{pol-sym}
\end{figure*} 
\begin{figure} 
\centerline{\hspace*{0.8cm}
\includegraphics[width=6.5cm]{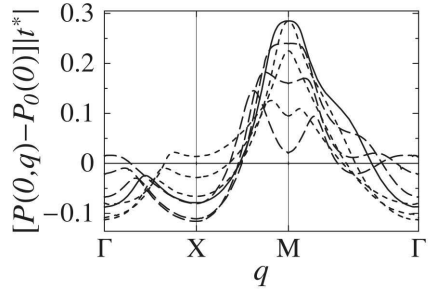}
}
\caption[5]{ 
$[P(0,{\bm q})-P_0(0)]|t^*|$ of the asymmetric model:
$k_{\rm B}T/|t^*|=\gamma/|t^*|=0.1$.
A solid line shows a result for $n \rightarrow 1$, dashed lines results
of electron doping cases such as $n=1.1$, 1.2 and 1.3, and  broken
lines  results for hole doping cases such as $n=0.9$, 0.8,  and
0.7. 
 }
\label{pol-asy}
\end{figure} 

Figures~\ref{pol-sym} and \ref{pol-asy} show $[P(0,{\bm q})-P_0(0)]$ of
the symmetric and asymmetric models. The polarization function at
${\bm q}=(\pm\pi/a,\pm\pi/a)$ is relatively larger in electron-doping
cases than it is in hole-doping cases.

\subsection{Reduction of $T_{\rm N}$ 
by the mode-mode coupling term}
\label{SecMode}
Following previous papers, \cite{kawabata,SCR,Miyake,OhkawaModeMode} 
we consider
mode-mode coupling terms linear  in intersite spin fluctuations
$F_{\rm s} ({\rm i} \omega_l, {\bm q}) $ given by eq.~(\ref{EqF}): 
\begin{equation}
\Lambda ({\rm i} \omega_l, {\bm q}) =
\Lambda_{\rm L }({\rm i} \omega_l)
+ \Lambda_{\rm s} ({\rm i} \omega_l, {\bm q})
+ \Lambda_{\rm v} ({\rm i} \omega_l, {\bm q}) .
\end{equation}
The first term $\Lambda_{\rm L} ({\rm i} \omega_l)$ is a {\it local} mode-mode
coupling term, which includes a single {\it local} four-point vertex
function, as is  shown in Fig.~4 of ref.~\citeonline{OhkawaModeMode}.
Both of  $\Lambda_{\rm s} ({\rm i} \omega_l, {\bm q})$ and
$\Lambda_{\rm v} ({\rm i} \omega_l, {\bm q})$ are {\it intersite} mode-mode coupling
terms, which include a single {\it intersite} four-point vertex function;
a single $F_{\rm s}({\rm i} \omega_l{\bm q})$
appears as the selfenergy correction to the single-particle Green
function in $\Lambda_{\rm s} ({\rm i} \omega_l, {\bm q})$
while it appears as a vertex correction to 
the polarization function  in
$\Lambda_{\rm v} ({\rm i} \omega_l, {\bm q})$, as are shown in Figs.~3(a) and 3(b),
respectively,  of ref.~\citeonline{OhkawaModeMode}.
Their static components are given by 
\begin{equation}\label{Lambda_L}
\Lambda_{\rm L}(0)
= \frac{5}{2\tilde{\chi}_{\rm s}(0)}\frac{k_{\rm B} T}{N}
\sum_{\omega_l{\bm q}} F_{\rm s}({\rm i} \omega_l,{\bm q}) ,
\end{equation}
\begin{equation}\label{Lambda_s}
\Lambda_{\rm s}(0,{\bm q}) =
\frac{3}{\tilde{\chi}_{\rm s}(0)} 
\frac{k_{\rm B}T}{N} \sum_{\omega_l, {\bm p}}
B_{\rm s}({\rm i} \omega_l,{\bm p};{\bm q})
%
\left[
F_{\rm s} ({\rm i} \omega_l, {\bm q}) 
- \frac1{4}J({\bm q}) \tilde{\chi}_{\rm s}^2 ({\rm i} \omega_l)
\right],
\end{equation}
\begin{equation}\label{Lambda_v}
\Lambda_{\rm v}(0,{\bm q}) =
-\frac{1}{2\tilde{\chi}_{\rm s}(0)} 
\frac{k_{\rm B}T}{N} \sum_{\omega_l, {\bm p}}
B_{\rm v}({\rm i} \omega_l,{\bm p};{\bm q})
F_{\rm s}({\rm i} \omega_l, {\bm q}) ,
\end{equation}
with 
\begin{eqnarray}\label{EqBs}
B_{\rm s}({\rm i} \omega_l,{\bm p};{\bm q})  &=& 
\frac{4\tilde{W}_{\rm s}^4}{\tilde{\chi}_{\rm s}^3(0)}
k_{\rm B}T \sum_{\varepsilon_n} \Biggl\{
\frac1{N} \sum_{\bm k}
g_\sigma^{(0)}({\rm i} \varepsilon_n,{\bm k}-{\bm q})
%
\left[g_\sigma^{(0)}({\rm i} \varepsilon_n,{\bm k})\right]^2
g_\sigma^{(0)}({\rm i} \varepsilon_n+{\rm i} \omega_l,{\bm k}+{\bm p})
\nonumber \\ && 
- \left[r_\sigma^{(0)}({\rm i} \varepsilon_n)\right]^3
r_\sigma^{(0)}({\rm i} \varepsilon_n+{\rm i} \omega_l)
\Biggr\} ,
\end{eqnarray}
\begin{eqnarray}\label{EqBv}
B_{\rm v}({\rm i} \omega_l,{\bm p};{\bm q})  &=& 
\frac{4\tilde{W}_{\rm s}^4}{\tilde{\chi}_{\rm s}^3(0)}
k_{\rm B}T \sum_{\varepsilon_n} \Biggl\{
\frac1{N} \sum_{\bm k}
g_\sigma^{(0)}({\rm i} \varepsilon_n,{\bm k}+{\bm q})
%
g_\sigma^{(0)}({\rm i} \varepsilon_n,{\bm k})
g_\sigma^{(0)}({\rm i} \varepsilon_n+{\rm i} \omega_l,{\bm k}+{\bm q}+{\bm p})
\nonumber \\  && \times 
g_\sigma^{(0)}({\rm i} \varepsilon_n+{\rm i} \omega_l,{\bm k}+{\bm p})
%
- \left[r_\sigma^{(0)}({\rm i} \varepsilon_n)\right]^2 \!
\left[r_\sigma^{(0)}({\rm i} \varepsilon_n \!+\! {\rm i} \omega_l)\right]^2
\Biggr\} ,
\end{eqnarray}
with
%
$r_\sigma^{(0)}({\rm i} \varepsilon_n) = 
(1/N) \sum_{\bm k} g_\sigma^{(0)}({\rm i} \varepsilon_n, {\bm k})$.
%
Because the selfenergy
correction linear in $J({\bm q})$ is considered in
\S~\ref{SecFock}, 
$(1/4)J({\bm q})\tilde{\chi}_{\rm s}^2({\rm i} \omega_l)$ is subtracted
in eq.~(\ref{Lambda_s}).

In this paper, weak three dimensionality  in spin fluctuations is
phenomenologically included.  Because
$J({\bm q})$ has its maximum value at 
${\bm q}= (\pm \pi/a, \pm\pi/a)$ and the nesting vector of the
Fermi surface in two
dimensions is close to 
${\bm q}=(\pm \pi/a, \pm \pi/a)$ for almost half filling, we assume
that the ordering wave number in three dimensions is 
%
${\bm Q} = (\pm \pi/a, \pm \pi/a, \pm Q_z)$, 
with $Q_z$ depending on interlayer exchange interactions. 
On the phase boundary between paramagnetic and antiferromagnetic
phases, where eq.~(\ref{EqAFCondition}) is satisfied, the
inverse of the susceptibility is expanded around ${\bm Q}$ and
for small $|\omega_l|$ in such a way that
\begin{equation}
\left[1/\chi_{\rm s}({\rm i} \omega_l,{\bm Q} + {\bm q})\right]_{T=T_{\rm N}} =
A({\bm q}) + \alpha_\omega |\omega_l|  + \cdots,
\end{equation}
with
\begin{equation}\label{EqAQ}
A({\bm q}) = \frac1{4}A_\parallel ({\bm q}_\parallel a)^2 
+ \frac1{4} A_z \left[( q_z - Q_z) c\right]^2 .
\end{equation}
Here, $c$ is the lattice constant along the $z$ axis. 
Because $\chi_{\rm s}({\rm i} \omega_l,{\bm Q} + {\bm q})$ diverges in
the limit of $|{\bm q}| \rightarrow 0$ and $\omega_l \rightarrow 0$
on the phase boundary, 
\begin{eqnarray}
B_{\rm s}(0,-{\bm Q};{\bm Q}) &=& 
B_{\rm v}(0,-{\bm Q};{\bm Q})
\nonumber \\ &=& 
\frac{4\tilde{W}_{\rm s}^4}{\tilde{\chi}_{\rm s}^3}
k_{\rm B} T \sum_{\varepsilon_n} \Biggl\{
\frac1{N}\! \sum_{\bm k}
\left[g_\sigma^{(0)}({\rm i} \varepsilon_n,{\bm k}\!-\!{\bm Q})\right]^2
%
\left[g_\sigma^{(0)}({\rm i} \varepsilon_n,{\bm k}) \right]^2
\!-\!\left[r_\sigma^{(0)} ({\rm i} \varepsilon_n)\right]^4
\Biggr\} , \hspace{1.1cm}
\end{eqnarray}
can be approximately used for $B_{\rm s}({\rm i} \omega_l,{\bm p};{\bm q})$ in
eq.~(\ref{Lambda_s}) and 
$B_{\rm v}({\rm i} \omega_l,{\bm p};{\bm q})$ in eq.~(\ref{Lambda_v}).   Then,  it
follows that
\begin{equation}
\Lambda (0,{\bm Q}) =
\frac{5}{2\tilde{\chi}_{\rm s}(0)} (1 + C_F - \tilde{C}_{\rm L}) \Phi ,
\end{equation} 
with
\begin{eqnarray}\label{EqCF}
C_F \! &=& \!
\frac{8W_s^4}{\tilde{\chi}_{\rm s}^3(0)}
\frac{1}{N} \sum_{\bm k}
\Biggl\{ \!
\frac{f_{\gamma}(\xi({\bm k} \!+\! {\bm Q}) \!-\! \mu^*)
\!-\! f_{\gamma}(\xi({\bm k}) \!-\! \mu^*)}
{\xi({\bm k})-\xi({\bm k}+{\bm Q})}
%
+ \frac1{2}
\Bigl[ f_{\gamma}^\prime(\xi({\bm k}) \!-\! \mu^*)
\!+\! f_{\gamma}^\prime(\xi({\bm k} \!+\! {\bm Q}) \!-\! \mu^*)\Bigr] \!\!
\Biggr\}
\nonumber\\ && \hspace*{5cm} \times 
\frac1{[\xi({\bm k})-\xi({\bm k}+{\bm Q})]^2} ,
\end{eqnarray}
\begin{equation}\label{EqCL}
\tilde{C}_{\rm L} = \frac{16 \tilde{W}_{\rm s}^4}{\tilde{\chi}_{\rm s}^3(0)}
\!\int \!\!d\varepsilon \Bigl[
\bar{\rho}(x)\bar{\rho}_2^3(\varepsilon) -
\pi^2\bar{\rho}^3(\varepsilon)\bar{\rho}_2(\varepsilon)
\Bigr] f_\gamma(\varepsilon-\mu^*) ,
\end{equation}
\begin{eqnarray}\label{EqPhi}
\Phi  &\equiv& 
\frac{k_{\rm B}T}{N} \sum_{\omega_l} \hspace{-5pt}~^{\prime} 
\sum_{|{\bm q}|\le q_ {\rm c}} \sum_{q_z}
\frac1{A({\bm q}) + \alpha_\omega |\omega_l|}
\nonumber \\ &=&
\frac{2 c}{\pi^3 A_\parallel} \hspace{-2pt}
\int_0^{\pi/c} \hspace{-8pt} dq_z  \hspace{-2pt}
\int_0^{\omega_ {\rm c}} \hspace{-8pt} d\omega 
\left[ n(\omega) + \frac1{2} \right] \!
%
\Biggl\{\tan^{-1}\!
\left[ \frac{A_\parallel (q_ {\rm c} a)^2 + A_z (q_zc)^2}
{4\alpha_\omega\omega}\right] 
%
- \tan^{-1}\!\left[ \frac{A_z (q_zc)^2}
{4\alpha_\omega\omega}\right]
\Biggr\} . \nonumber \\ &&
\end{eqnarray}
In eq.~(\ref{EqCF}), $f_{\gamma}^\prime (\varepsilon)$  is the
derivative of $f_\gamma(\varepsilon)$ defined by eq.~(\ref{EqfGamma}).
 In eq.~(\ref{EqCL}), $\rho_{\gamma\rightarrow 0}(\varepsilon)$ is 
denoted by $\bar{\rho}(\varepsilon)$, and 
\begin{equation}
\bar{\rho}_2(\varepsilon) = \mbox{Vp}\! \int d\varepsilon^\prime 
\frac{\bar{\rho}(\varepsilon^\prime)}
{\varepsilon-\varepsilon^\prime} .
\end{equation}
In eq.~(\ref{EqPhi}), the summation is restricted to
$|\omega| \le \omega_ {\rm c}$ and 
$|{\bm q}_\parallel| \le q_ {\rm c} $. 
We assume that $q_ {\rm c}=\pi/3a$ and $\omega_ {\rm c}$ is given by a larger one of
$8[t^*|$ and  $|J|$.

As was shown in the previous paper, \cite{OhkawaPseudogap}   
$\alpha_\omega\simeq 1$ for $T/T_{\rm K} \alt 1$ and $\gamma/k_{\rm B}T_{\rm K} \alt 1$.
Physical properties for $T/T_{\rm K} \gg 1$ or $\gamma/k_{\rm B}T_{\rm K} \gg 1$
scarcely depends on $\alpha_\omega$. Then, we assume  $\alpha_\omega
=1$ for any $T$ and $\gamma$ in this paper.

\subsection{Almost symmetric $T_{\rm N}$ between $\delta>0$ and
$\delta<0$} 
\label{SecAlmost}

\begin{figure} 
\centerline{\hspace*{1cm}
\includegraphics[width=7.5cm]{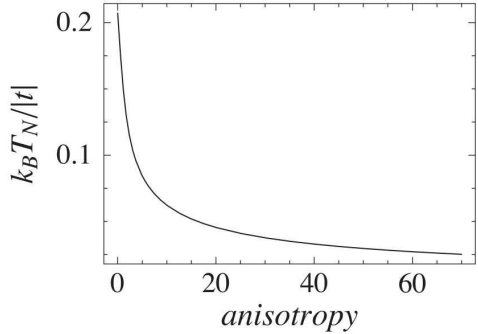}
}
\caption[6]{ 
$T_{\rm N}$ determined by eq.~(\ref{EqAFCondition2}) for $J/|t| =-0.3$ 
and $\delta=0$ 
as a function of the exponent $x$ of the anisotropy factor  defined by
$|A_z/A_\parallel| = 10^{-x}$.
 }
\label{TN-asym}
\end{figure}

The Kondo temperature $T_{\rm K}$ is low in the limit of $\delta\rightarrow 0$
for nonzero $T$ or $\gamma$.
When $T/T_{\rm K} \gg 1$ or $\gamma/k_{\rm B}T_{\rm K} \gg 1$, the instability condition
(\ref{EqAFCondition}) becomes simple:
\begin{equation}\label{EqAFCondition2}
\frac1{\tilde{\chi}_{\rm s}(0)} + \frac{5\Phi}{2\tilde{\chi}_{\rm s}(0)} 
- \frac1{4}J({\bm Q}) =0 .
\end{equation}
When we ignore $\Lambda_{\rm L}(0)=5\Phi/2\tilde{\chi}_{\rm s}(0)$ and we
assume $T_{\rm K} =0$,  eq.~(\ref{EqAFCondition2}) gives 
$T_{\rm N} = (1/4)J({\bm Q})/k_{\rm B}=|J|/k_{\rm B}$. This $T_{\rm N}$ is nothing but $T_{\rm N}$ in
the mean-field approximation for the Heisenberg model, to which the
$t$-$J$ model with the half filling is reduced.  On the
other hand, $\Phi$ diverges at non zero temperatures for $A_z=0$; no
magnetic instability occurs at nonzero temperature in complete two
dimensions. \cite{Mermin}  The reduction of
$T_{\rm N}$ by  $\Lambda_{\rm L}(0)$ or by quasi-two dimensional thermal spin
fluctuations depends on the anisotropy of 
$A_z/A_\parallel$. When only the
superexchange interaction  is considered,  $A_\parallel = |J|$.  Although 
$(1/4)J_{\rm Q}(0,{\bm Q}+{\bm q})$ also contribute to the $q$-quadratic term,
we assume $A_\parallel =|J|$ for the sake of simplicity. 
Figure \ref{TN-asym} show $T_{\rm N}$ as a function of the exponent $x$ of
$|A_z/A_\parallel| = 10^{-x}$ for the half filling $(\delta=0)$.
When we take $x=10$, for example, 
we obtain $T_{\rm N}\simeq 0.06 |t|/k_{\rm B}$ or 
$T_{\rm N}\simeq 0.2 |J|/k_{\rm B}$.
We assume this anisotropy factor  $x=10$ in the following part of this paper.

\begin{figure} 
\centerline{\hspace*{1.2cm}
\includegraphics[width=7.5cm]{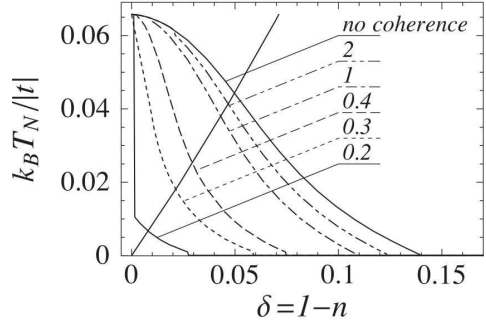}
}
\caption[7]{ 
$T_{\rm N}$ of the symmetric model as a function of $\delta= 1-n$.  From the
bottom, solid, dashed, broken, chain, and chain double-dashed lines
show $T_{\rm N}$ for $\gamma/|t|=0.2$, 0.3, 0.4, 1, and 2, respectively. The
topmost solid line shows $T_{\rm N}$ determined from
eq.~(\ref{EqAFCondition2}) for comparison.  The solid line with a
positive slope shows $1/\tilde{\phi}_\gamma$.  Antiferromagnetic states
whose $T_{\rm N}$ are much above and below this line are
characterized as local-moment ones and
itinerant-electron ones, respectively,
according to discussion in \S~\ref{SecDiscussion}.
 }
\label{Neel-TN}
\end{figure} 
\begin{figure} 
\centerline{\hspace*{1cm}
\includegraphics[width=7.5cm]{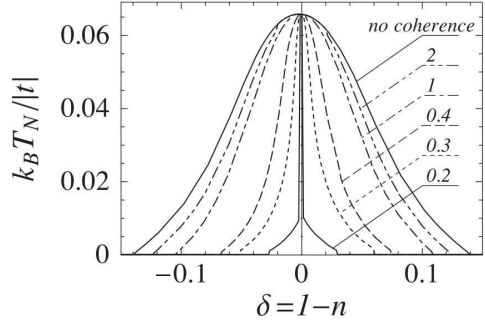}
}
\caption[8]{ 
$T_{\rm N}$ of the asymmetric model as a function of $\delta$.  See also the
caption of Fig.~\ref{Neel-TN}. Note that $T_{\rm N}$ is almost symmetric between
$\delta<0$ and $\delta>0$.
 }
\label{Neel-TN-asym}
\end{figure} 

Figures~\ref{Neel-TN} and \ref{Neel-TN-asym} show $T_{\rm N}$ of the symmetric and
asymmetric models as a function of  $\delta=1-n$ for various $\gamma$.
The antiferromagnetic phase becomes narrower as $\gamma$ decrease. As long as
$\gamma$ is almost symmetric between
$\delta>0$ and $\delta<0$, $T_{\rm N}$ is also almost symmetric between them.

When $\gamma\gg W^*$ or $k_{\rm B}T\gg W^*$, $T_{\rm N}$ is determined by the
competition between the superexchange interaction $J$ and two quenching effects,
$1/\tilde{\chi}_{\rm s}(0)$ and $\Lambda_{\rm L}(0)$.  The reduction of $T_{\rm N}$ for small
$|\delta|$ is mainly due to $\Lambda_{\rm L}(0)$. On the other hand, the critical
concentration $|\delta_ {\rm c}|$ below which antiferromagnetic ordering appears at
$T=0$~K is determined by the competition between $J$ and $k_{\rm B} T_{\rm K}$,
because thermal spin fluctuations vanish at
$T=0$~K. Because the Kondo effect is rather weak, $|\delta_ {\rm c}|$ is as large as
$|\delta_ {\rm c}|\simeq 0.14$, as is shown in Figs.~\ref{Neel-TN} and
\ref{Neel-TN-asym}.

When $\gamma\alt W^*$ and $k_{\rm B}T\alt  W^*$, well-defined quasiparticles can play
roles in both of the enhancement and suppression of $T_{\rm N}$. Antiferromagnetism
is enhanced by $J_{\rm Q}({\rm i} \varepsilon_n, {\bm q})$,  in which the nesting of the Fermi
surface can play a significant role. However, the difference of the Fermi surfaces
cannot give any significant  asymmetry of $T_{\rm N}$.  The reduction of $T_{\rm
N}$ by the Kondo effect is large.   The intersite mode-mode coupling term,
$\Lambda_{\rm s}(0,{\bm Q})+\Lambda_{\rm v}(0,{\bm Q})$, also suppresses $T_{\rm N}$ in
addition to the local terms, $k_{\rm B}T_{\rm K}$ and
$\Lambda_{\rm L}(0)$.
The quenching effects overcome the enhancement effect,
and $T_{\rm N}$ decrease with decreasing $\gamma$.

\section{Discussion}
\label{SecDiscussion}

One of the most serious assumptions in this paper is that the homogeneous
life-time width $\gamma$ of quasiparticles,
which is phenomenologically introduced in this paper,
can be defined. This is relevant when
there are many weak impurities 
and their distribution is totally random.
When disorder is so small that quasiparticles are well defined,
electrons behave as Landau's Fermi liquid.
The assumption is obviously relevant
 in such a case.  On the other hand,
when disorder is so large in the vicinity of the Mott transition or
crossover  that quasiparticles are never well-defined,
electrons behave as localized magnetic moments.
Because $\gamma$  plays no role, the theoretical framework
of this paper must also be relevant in such a case.
Then, an issues  is  whether  the assumption is relevant  or irrelevant
in the crossover regime between the two regimes,
for example, in the regime where the Anderson localization 
of  quasiparticles occurs.
Because the homogeneous lifetime width can be defined
in the so called weakly localized regime of
the Anderson localization, \cite{weaklocalization} we expect that
the results of this paper apply to such a regime.
The assumption is irrelevant  when scattering
potentials of dopants are strong and the concentration of dopants is
rather small. 
The strong Anderson localization, 
where the localization length is short, 
is out of scope in this paper.  
The crossover between the Mott-Hubbard insulator and the Anderson insulator,
which is examined in ref.~\citeonline{byczuk}, 
is also out of scope in this paper.

One may suspect that  the theoretical framework of this paper
cannot apply to the limit of
$\delta \rightarrow 0$ or to vanishingly small $|\delta|$.
According to the combination of Gutzwiller 's theory \cite{Gutzwiller}
and the Fermi-liquid theory, \cite{Luttinger1,Luttinger2}
the bandwidth of quasiparticles is vanishingly small but is still nonzero such
as $W^* \simeq 8|\delta t|$.
The Fermi liquid state is formed, at least,  at $T \ll 8|\delta t|/k_{\rm B}$; 
the spectral weight of quasiparticles is vanishingly small.
 It is obvious that
when the renormalization of quasiparticles due to the Fock-type term
of the superexchange interaction is taken into account  the bandwidth
becomes as large as $|J|$,  at least, at $T \ll 8|\delta t|/k_{\rm B}$.
We show in this paper 
that these quasiparticles are stable not only for
$k_{\rm B}T \ll 8|\delta t|$ and $\gamma \ll 8|\delta t|$
but also for  $k_{\rm B}T \ll |J|$ and  $\gamma \ll |J|$.
The existence of elementary excitations or
quasiparticles whose bandwidth is of the order of $|J|$
are implied by several  studies on the $t$-$J$ model on small lattices.
\cite{tohyama-t-J,moreo,wei-cheng}
Then,  we can argue that when the Hilbert space
is restricted within paramagnetic phases
no divergence of the effective mass of quasiparticles or 
no M-I transition can  occur except for
$J\rightarrow 0$, $\delta\rightarrow 0$, and $T\rightarrow 0$~K.
What can occur for nonzero $J$ is 
a M-I crossover rather than a M-I transition,  
as is shown  in Figs.~\ref{t-star} and \ref{t-star-asym}.

The coexistence phase of
a metal and a Mott-Hubbard insulator is obtained for 
the Hubbard model with the just half filling and finite onsite $U$. \cite{byczuk}
It is interesting to examine whether or not the coexistence phase
remains  when the renormalization
of quasiparticles by an intersite exchange interaction such as the superexchange
interaction  is included.
It is also interesting to examine whether or not the coexistence phase 
remains for non-half fillings.

Quasiparticles are well-defined 
for $T\alt T_{\rm K}$ and $\gamma\alt k_{\rm B}T_{\rm K}$.
According to refs.~\citeonline{ohkawaCW} and  \citeonline{miyai},
the exchange interaction $J_{\rm Q}(+{\rm i} 0, {\bm q})$, which arises from the
virtual exchange of pair excitations of quasiparticles, shows an almost
$T$-linear dependence consistent with the Curie-Weiss law
at $T\alt T_{\rm K}$ in a small region of  ${\bm q} \simeq {\bm Q}$,
with ${\bm Q}$ being the nesting  wavenumber, for
itinerant-electron antiferromagnets where the nesting of the Fermi surface 
is significant. \cite{ComCW} 
On the other hand, the $T$-linear dependence of
$1/\tilde{\chi}_{\rm s}(0)$  at $T\gg T_{\rm K}$ is responsible for the Curie-Weiss
law of local-moment magnets or insulating magnets. Then, magnetism with
$T_{\rm N} \gg T_{\rm K}$ is characterized as local-moment one, 
while magnetism with 
$T_{\rm N}\alt T_{\rm K}$ is characterized as itinerant-electron one. 
Local-moment magnetism appears for almost half fillings, and
itinerant-electron magnetism appears for fillings substantially away from the half filling.

Because no order is possible at non-zero temperatures in two dimensions,
\cite{Mermin}
it is likely that $T_{\rm N}$ is substantially reduced by critical thermal  fluctuations
 in quasi-two dimensions.
As is examined in \S~\ref{SecAlmost}, the reduction of $T_{\rm N}$ is large;
$T_{\rm N}\simeq 0.2 |J|/k_{\rm B}$ for the anisotropy of
$|A_z/A_\parallel| =10^{-10}$.  This explains observed $T_{\rm N} \simeq 300$~K for
cuprate oxides, when we take $|J|\simeq 0.15$~eV.   The exponent $10$  seems to be
little larger than actual ones.  We should consider the reduction of $T_{\rm N}$
by thermal critical fluctuations more properly than we do in this paper.

The nesting of the Fermi surface becomes less sharp as $\gamma$ becomes
larger, as is shown in Figs.~\ref{pol-sym} and \ref{pol-asy}.
One may argue that antiferromagnetism is weaken by disorder 
in the itinerant-electron regime where the nesting of the Fermi surface
plays a crucial role.
However, this effect is overcome by the reduction of $T_{\rm K}$
by nonzero $\gamma$.
The renormalization  of the quasiparticle bandwidth 
by the superexchange interaction  becomes
small with increasing $\gamma$ so that  
the Kondo temperature $k_{\rm B}T_{\rm K}$ becomes small. 
On the other hand, the superexchange interaction
is not reduced by nonzero $\gamma$. 
Then, antiferromagnetism is strengthen with increasing $\gamma$ even in the
itinerant-electron region.

An antiferromagnetic states appears in a narrow range of $0\le
|\delta|\alt 0.02$--$0.05$ in hole-doped cuprates oxides $(\delta>0)$, while it
appears in a wide range of $0\le|\delta| \alt 0.13$--$0.15$ in
electron-doped cuprate oxides $(\delta<0)$.  
Tohyama and Maekawa \cite{tohyama} argued that the asymmetry between 
hole-doped and electron-doped oxides must arises
from the difference of the Fermi surfaces, and that  the
$t$-$t^\prime$-$J$  model should be used.
They showed that  the intensity of spin excitations is relatively
stronger in electron doping cases than it is in hole doping cases. 
According to Fig.~\ref{pol-asy}, the polarization function at
${\bm q}=(\pm\pi/2a,\pm\pi/2a)$  is relatively larger in electron doping
cases than it is in hole doping cases. This asymmetry is consistent with
that of spin excitations studied by Tohyama and Maekawa.   However, the
difference of the Fermi surfaces cannot explain the asymmetry of $T_{\rm N}$, as
is shown in Fig.~\ref{Neel-TN-asym}.

The condensation energy at $T=0$~K of the asymmetric model is also quite
asymmetric; \cite{yokoyama} it is consistent with the asymmetry of 
spin excitations and the polarization function discussed
above. On the other hand, $T_{\rm N}$ is significantly reduced by quasi-two
dimensional spin fluctuations as well as local spin fluctuations of the
Kondo effect. This large reduction of $T_{\rm N}$ arises from the
renormalization of normal states; not only the N\'{e}el states but also
paramagnetic states just above $T_{\rm N}$ are largely renormalized.  It
is plausible that the asymmetry of the condensation energy of
paramagnetic states just above $T_{\rm N}$ is similar to that of the N\'{e}el
states at $T=0$~K. It is interesting to examine by comparing the
condensation energy of the N\'{e}el states and that of paramagnetic
states just above $T_{\rm N}$ whether $T_{\rm N}$ is actually almost symmetric as is
shown in this paper. 

Electrical resistivities of electron-doped cuprates are relatively larger
than those of hole-doped cuprates are. \cite{tokura}   Then, we can argue
that disorder must be relatively larger in electron-doped cuprates than
it is in hole-doped cuprates, and that the asymmetry of $T_{\rm N}$  can arise,
at least partly, 
from the difference of disorder between them.
It is interesting to examine whether or not the
symmetric behavior of $T_{\rm N}$ can
be restored by preparing hole-doped and electron-doped cuprates with
similar degree of disorder to each other.

When nonmagnetic impurities of Zn ions are introduced on CuO$_2$ planes, 
antiferromagnetic moments appear
in the neighborhood of Zn ions. \cite{CuprateZn} 
Because the configuration of Cu ions is $(3d)^9$ and that of Zn ions
is $(3d)^{10}$,
the phase shift at the chemical potential due to scatterings 
by Zn ions must be about $\pi/2$ according to the Friedel sum rule. Then,
scatterings by Zn ions must be very strong.
One of the plausible scenarios is that
the renormalization of $k_{\rm B}T_{\rm K}$ by the Fock-type term of
the superexchange interaction is small in the neighborhood of Zn ions
because of strong scatterings,
that is, the quenching of magnetic moments by local quantum spin fluctuations
 are weak there so that
antiferromagnetism is enhanced there.
It is interesting to develop a microscopic theory that can treat
such an inhomogeneous effect.

\begin{figure} 
\centerline{\hspace*{1.2cm}
\includegraphics[width=7.5cm]{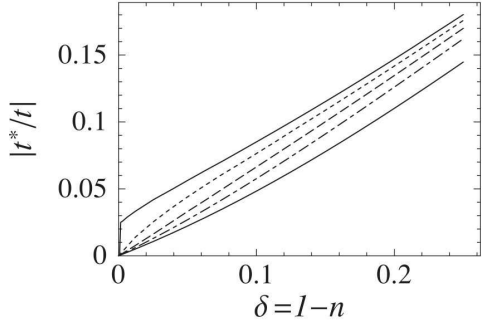}
}
\caption[10]{ 
$t^*$ of the symmetric model as a function of $\delta$ for various $\gamma$
and $k_{\rm  B}T/|t|=0.02$; $\tilde{W}_{\rm s}\simeq 1$ or $r=0.5$ is assumed instead of
$\tilde{W}_{\rm s}\simeq 2$ or  $r=1$ (See text). From the top, solid, 
dashed, broken, and chain line show $t^*$ for
$\gamma/|t|=0.04$,  0.1, 0.2, and 0.4, respectively.  For the sake of
comparison, $1/\tilde{\phi}_\gamma$ is also shown by a bottom solid line. 
}
\label{tstar-1}
\end{figure} 
\begin{figure} 
\centerline{\hspace*{1.2cm}
\includegraphics[width=7.5cm]{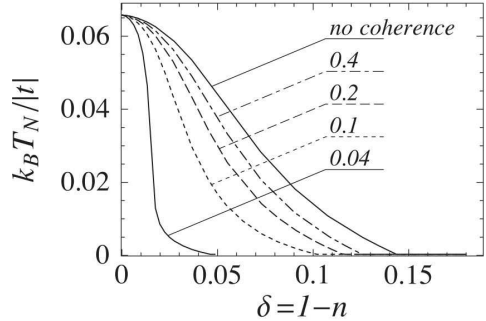}
}
\caption[11]{ 
$T_{\rm N}$ of the symmetric model as a function of $\delta$ for various
$\gamma$; $\tilde{W}_{\rm s}\simeq 1$ or  $r=0.5$ is assumed instead of
$\tilde{W}_{\rm s}\simeq 2$ or $r=1$ (See text). From the bottom, solid,
dashed, broken, and chain lines show $T_{\rm N}$ for $\gamma/|t|=0.04$,
0.1, 0.2, and 0.4, respectively. For comparison,  $T_{\rm N}$
determined from eq.~(\ref{EqAFCondition2}) is shown by a topmost
solid line.    
 }
\label{Neel-TN3}
\end{figure} 

The three-point vertex function $\tilde{\phi}_{\rm s}$ and the
mass-renormalization factor $\tilde{\phi}_\gamma$ are those in SSA;
$\tilde{W}_{\rm s} \equiv \tilde{\phi}_{\rm s}/\tilde{\phi}_\gamma \simeq 2$
as is discussed in  \S~\ref{SecRenomalizedSSA}.
When we take $\tilde{W}_{\rm s}\simeq 2$, on the other hand, 
the superexchange interaction  constants as small as $|J|= 0.02\mbox{-}0.03$~eV,
which are much smaller than experimental ones 
$|J|= 0.10\mbox{-}0.15$~eV, give 
$T_ {\rm c}$ as high as $T_ {\rm c}=50\mbox{-}150$~K, 
\cite{Ohkawa87SC-1,Ohkawa87SC-2} which are
as high as observed $T_ {\rm c}$. This implies that
the renormalization of the
vertex function and the mass renormalization factor
by intersite effects such as
antiferromagnetic and superconducting fluctuations
should be included.  Taking it into account, 
we treat $\tilde{W}_{\rm s}$ as another phenomenological
parameter following the previous paper,\cite{OhkawaPseudogap} where
$\tilde{W}_{\rm s}=0.7\mbox{--}1$ is used instead of $\tilde{W}_{\rm s}\simeq 2$ in
order to explain observed $T_ {\rm c}$ and observed
coefficients of $T$-linear resistivities
when $|J|= 0.10\mbox{-}0.15$~eV are used. 
Then, we replace $\tilde{W}_{\rm s}$ by
$r\tilde{W}_{\rm s}$, with $r$ a numerical constant smaller
than unity; 
eq.~(\ref{Eqt*}) is replaced by
\begin{equation}\label{EqReplace}
2 t^* = r \frac{2t}{\tilde{\phi}_\gamma} 
- \frac{3}{4}\left(r\tilde{W}_{\rm s}\right)^2 J \Xi ,
\end{equation}
and eqs.~(\ref{EqTK2}) and (\ref{EqTKdef}) are replaced by those with
$r\tilde{W}_{\rm s}$ instead of $\tilde{W}_{\rm s}$. \cite{comment-Ws} 
Figures~\ref{tstar-1} and \ref{Neel-TN3} show 
$t^*$ and $T_{\rm N}$, respectively, 
of the symmetric model as a function of $\delta$ for $r=0.5$ and various
$\gamma$.  
The antiferromagnetic region extends with decreasing $r$. 

An antiferromagnetic state in the range of $0\le\delta\alt 0.02$ of
hole-doped La$_{2-\delta}$M$_\delta$CuO$_4$ (M= Sr or Ba) is
characterized as a local-moment one. The so called spin-glass or
Kumagai's phase\cite{Kumagai} appears in the range of
$0.02\alt\delta\alt 0.05$. 
We characterize the tail part of the solid line
in Fig.~\ref{Neel-TN3} as a critical line to the Kumagai phase; \cite{g-d}
$\gamma/|t|\simeq 0.04$ are needed to reproduce Kumagai's phase.
According to Fig.~\ref{tstar-1},
$|t^*/t|\simeq 0.05$ for $\delta=0.04$. Then, we can argue
that  $k_{\rm F} l \simeq 2k_{\rm  B}T_{\rm K}/\gamma \simeq 4|t^*|/\gamma$, 
with $k_{\rm F}$ the
Fermi wave number and
$l$ the mean free path, must be 4--8 in Kumagai's
phase.
 According to Fig.~\ref{pol-sym}(b), the
nesting of the Fermi surface is  substantial at least for $\gamma/|t^*|
\alt 0.3$; the nesting cannot be ignored for $\gamma/|t^*| \alt 1$.
Kumagai's phase must be a spin density wave (SDW) state in a disordered
system rather than a spin glass. The divergence\cite{Kumagai} of the
nuclear quadrupole relaxation (NQR) rate at $T_{\rm N}$  supports this
characterization.

\section{Conclusion}
\label{SecConclusion}

The  $t$-$J$ model with $J/|t|=-0.3$  on a quasi-two dimensional lattice
is studied.
First, an {\it  unperturbed} state is constructed in SSA or DMFT. 
The Kondo temperature $T_{\rm K}$ or $k_{\rm B}T_{\rm K}$ is defined as the energy scale of
local quantum spin fluctuations. 
 The unperturbed state is a normal degenerate or almost degenerate 
 Fermi liquid at $T \alt T_{\rm K}$
and is a non-degenerate Fermi liquid at  $T \gg T_{\rm K}$.
No metal-insulator transition is possible
within paramagnetic phases;
what occurs is a metal-insulator crossover.
The bandwidth $W^*$ of  quasiparticles, 
which  is about $W^*\simeq4 k_{\rm B}T_{\rm K}$,
is renormalized by the Fock-type 
selfenergy due to the superexchange interaction $J$, so that that
$W^*$ is larger for lower $T$ and smaller  $\gamma$,
with $\gamma$ being the phenomenologically introduced 
lifetime width  of quasiparticles.
 The bandwidth $W^*$ is approximately given by
the sum of terms of $O(|\delta t|)$ and  $O(|J|)$, 
with  $\delta$ being the concentration of dopants
added to the just half-filled system; 
the term of $O(|\delta t|)$
is consistent with Gutzwiller's theory, and the coefficient of 
the term of  $O(|J|)$ is  $O(1)$ for
$k_{\rm B}T/|W^*| \alt 1$ and $\gamma/ |W^*| \alt 1$ and
is vanishingly small 
for $k_{\rm B}T/|W^*| \gg 1$ or $\gamma/ |W^*| \gg 1$.

Next, 
antiferromagnetic instability of the unperturbed state is examined  perturbatively
in terms of intersite exchange interactions.
This perturbation theory can treat both of local-moment magnetism
and itinerant-electron magnetism.
When $T_{\rm N} \gg T_{\rm K}$, with $T_{\rm N}$ the N\'{e}el temperature,
 or local thermal spin fluctuations are dominant over 
local quantum spin fluctuations at  $T_{\rm N}$, 
magnetism is characterized as
local-moment magnetism; it appear for almost half fillings of electrons.
In the opposite case where $T_{\rm N} \alt T_{\rm K}$ or
local quantum ones are dominant over 
local thermal ones at $T_{\rm N}$, magnetism is characterized as
itinerant-electron magnetism; it appears for fillings away from
the half filling.

The reduction of  $T_{\rm N}$ by
quasi-two dimensional thermal critical fluctuations is large
when the anisotropy of the exchange interaction constants is large.
For example, 
$k_{\rm B}T_{\rm N}/|J|\rightarrow 0$ in complete two dimensions, as is expected.
The N\'{e}el temperature
$T_{\rm N} \simeq 300$~K  observed in cuprates with no doping 
can be explained, when we assume $|J| = 0.10\mbox{-}0.15$~eV and 
$|J_z/J|\simeq 10^{-10}$, with
$J$ and $J_z$  the exchange interaction constants between nearest neighbors 
within a CuO$_2$ plane and between CuO$_2$ planes,
respectively.  

When $\gamma \gg W^*$ or $k_{\rm B}T \gg W^*$, 
$W^* = O(|\delta t|)$ or  $k_{\rm B}T_{\rm K} = O(|\delta t|)$ so that
the quenching of magnetic moments
 by local quantum spin fluctuations is rather weak.  
 Because thermal spin fluctuations vanish at
$T=0$~K, an antiferromagnetic state is stabilized  in a wide
range of $\delta$.  
When $\gamma\alt W^*$ and $k_{\rm B}T\alt W^*$, 
quasiparticles are well-defined. Because 
the nesting of the Fermi surface is substantial, 
the exchange interaction arising from the virtual exchange of pair excitations of
quasiparticles is also responsible for antiferromagnetic instability in addition to the
superexchange interaction. On the other hand, 
the quenching of magnetic moments
 by local quantum spin fluctuations is  strong. 
Not only quasi-two dimensional thermal
critical fluctuations but also 
the quenching of magnetic moments
by local quantum fluctuations make $T_{\rm N}$ substantially reduced
or they destroy antiferromagnetic ordering. 
An antiferromagnetic state is stabilized  only for small $|\delta|$;
the critical concentration $|\delta_ {\rm c}|$
below which antiferromagnetic ordering appears is smaller for
smaller $\gamma$. 
This result implies that  $T_{\rm N}$ is lower and
a magnetic phase is narrower in a cleaner system.

The difference or asymmetry of  disorder between
electron-doped and hole-doped cuprates 
must be, at least partly, responsible for that
of $T_{\rm N}$ and antiferromagnetic phases between them.
We characterize the so called Kumagai's phase as
a spin density wave (SDW) state in disordered system rather than
a spin glass. 

\begin{acknowledgments}
The author is thankful for discussion to M. Ido, M.
Oda, N. Momono, and K. Kumagai.  This work was supported by a Grant-in-Aid for
Scientific Research (C) Grant No.~13640342 from the Ministry of
Education, Cultures, Sports, Science and Technology of Japan. 
\end{acknowledgments}

\appendix

\section{Scattering Potential in Disordered Kondo Lattices}
\label{SecDisorder}

In disordered Kondo lattices, the mapped Anderson models are
different from site to site. The selfenergy for the
$j$th one is expanded in such a way that
$\tilde{\Sigma}_{j\sigma}(\varepsilon \!+\! {\rm i} 0) =
\tilde{\Sigma}_{j\sigma}(0) + \left(1-\tilde{\phi}_{j\gamma} \right)
\varepsilon + \cdots $.
When the energy dependence of the hybridization energy
$\Delta_j(\varepsilon)$ is ignored,
it follows according to Shiba \cite{shiba} that
\begin{equation}\label{EqSiteSigma}
E_{dj}+\tilde {\Sigma}_{j\sigma}(+{\rm i} 0) = 
\Delta_j(0) \tan \left[
\pi \left( \mbox{$\frac1{2}$} - n_{j\sigma} \right)\right] ,
\end{equation}
with $E_{dj}$ and
$n_{j\sigma}$ being the localized-electron level and the number of
electrons with spin $\sigma$, respectively. 
It follows from Eqs.~(\ref{EqPhiG}) and (\ref{EqPhiS}) that 
$\tilde{\phi}_{j\gamma} \simeq (\pi^2/8)/|1-n_j|$  and 
$\tilde{\phi}_{j{\rm s}} \simeq (\pi^2/4)/|1-n_j|$, with 
$n_j \equiv n_{j\uparrow} + n_{j\downarrow}$,
for almost half filling 
$n_j \simeq 1$.
We assume non-magnetic impurities so that
$n_{j\uparrow}=n_{j\downarrow}$.

We denote the average number of electrons and their mean-square deviation
by $n=\left<n_j\right>_{\rm dis}$ and 
$\Delta n^2=\left<(n_j-n)^2\right>_{\rm dis}$, respectively, where
$\left< \cdots \right>_{\rm dis}$ stands for an ensemble average over
disordered systems. 
We assume that there is no correlation between 
different sites:
$\left<(n_i-n)(n_j-n)\right>_{\rm dis}=0$ for $i\ne j$. 

We consider the site-dependent part of 
$E_{dj}+\tilde {\Sigma}_{j\sigma}(\varepsilon+{\rm i} 0)$ as a static
but energy-dependent 
scattering potential; it is approximately given by
\begin{equation}\label{EqScatt}
V_{j\sigma}(\varepsilon) =
- \left( \frac{\pi \Delta}{2}  
+ \frac{8}{\pi^2}\tilde{\phi}_\gamma^2 \varepsilon \right)
(n_{j} \!-\! n ) 
\frac{1 \!-\! n}{|1 \!-\! n|}+
\cdots .
\end{equation}
Here, the averages of $\tilde{\phi}_{j\gamma}$ and $\Delta_j(0)$
are simply dented by $\tilde{\phi}_\gamma$ and $\Delta$. When we treat
$V_{j\sigma}(\varepsilon)$ in the second-order SSA or the
Born approximation, the coherent part of the  ensemble averaged Green
function is given by
\begin{equation}
\left<G_\sigma^{(0)}(\varepsilon+{\rm i} 0,{\bm k}) \right>_{\rm dis} =
 \frac1{\tilde{\phi}_\gamma} \frac1
{\displaystyle \varepsilon  +\mu^* - \xi({\bm k}) - 
(1/\tilde{\phi}_\gamma) \Sigma_{\sigma}^{({\rm dis})} (\varepsilon+{\rm i} 0)  } ,
\end{equation}
with
\begin{equation}
\frac1{\tilde{\phi}_\gamma} \Sigma_{\sigma}^{({\rm dis})} 
(\varepsilon \!+\! {\rm i} 0) 
= - {\rm i} \frac{\pi}{\tilde{\phi}_\gamma^2}
\left[ \frac{\pi \Delta}{2} 
\!+\! \frac{8}{\pi^2}\tilde{\phi}_\gamma^2 \varepsilon
\right]^2  \!\!\Delta n^2 \rho_\gamma (\varepsilon) , 
\end{equation}
for $|\varepsilon| \alt k_{\rm B} T_{\rm K} $.

The bandwidth of quasiparticles is about $4 k_{\rm B}T_{\rm K}$; typical
lifetime width is as large as 
\begin{equation}
\frac1{\tilde{\phi}_\gamma} \mbox{Im}
\Sigma_{\sigma}^{({\rm dis})} (\pm k_{\rm B}T_{\rm K} +{\rm i} 0) 
\simeq - {\rm i} (64/\pi^3)
\tilde{\phi}_\gamma \Delta n^2  |t| . 
\end{equation}
The energy-independent term can be ignored because 
$\tilde{\phi}_\gamma\gg 1$ for almost half fillings. It is quite
likely that $\tilde{\phi}_\gamma \Delta n^2 =O (1)$ and 
$\gamma/|t|=O(1)$ for almost half fillings.

It is straightforward to extend the above argument to a system in the
presence of magnetic fields
and a system
with magnetic impurities. In such cases,
$n_{i\uparrow}-n_{i\downarrow}$ can be different from site to site; 
$\mbox{Im}\Sigma_{\sigma}^{({\rm dis})} (\varepsilon+{\rm i} 0)/\tilde{\phi}_\gamma$
can be large even for $\varepsilon=0$.




\begin{thebibliography}{}
\bibitem{ComGamma} 
Because of this simplification,
we can carry out analytically  summations over
Matsubara's frequencies in many equations of this paper.
\bibitem{ComCW} 
In itinerant-electron ferromagnets where the density of state has a sharp peak
around the chemical potential, the exchange interaction $J_{\rm Q}(+{\rm i} 0, {\bm q})$
shows an almost $T$-linear dependence
consistent with the Curie-Weiss law in a small region of ${\bm q}\simeq 0$
\cite{miyai}.  
In  itinerant-electron antiferromagnets
where the nesting of the Fermi surface is never significant and 
itinerant-electron ferromagnets where the density of states is almost
constant around the chemical potential, on the other hand,
the mode-mode coupling term shows an almost $T$-linear dependence
consistent with the Curie-Weiss law \cite{SCR}.
\bibitem{comment-Ws}
In eqs.~(\ref{EqTK2}) and (\ref{EqTKdef}), we should use a
multiplication factor slightly different from  $r$ because
$\tilde{\phi}_{\rm s}$ appearing there is a purely single-site property.  In
the first term of eq.~(\ref{EqReplace}), we should also use another
one slightly different from $r$  because no $\tilde{\phi}_{\rm s}$ appears
there. However, we  assume  the same $r$ in every equation for the
sake of simplicity; this corresponds to an approximation where
only the renormalization of the mass-enhancement factor is
considered.
\bibitem{g-d}
The dependence of $\gamma$ on $\delta$ should be included 
in order to explain more precisely the dependence of $T_{\rm N}$ on $\delta$.
\end{thebibliography}
\end{document}